\documentclass[a4paper,11pt]{article}
\usepackage[utf8]{inputenc}
\pdfoutput=1 

\usepackage{jcappub} 

\usepackage[T1]{fontenc} 
\usepackage{graphicx}
\usepackage[makeroom]{cancel}
\usepackage{makecell}
\usepackage{bm}
\usepackage{amssymb}
\usepackage{amsmath}
\usepackage{epsfig}
\usepackage{hyperref}
\usepackage{mathtools}
\usepackage{hhline}
\usepackage{multirow}
\usepackage[normalem]{ulem}
\usepackage{subfig}
\usepackage{tikz}
\usepackage{epstopdf}
\usetikzlibrary{snakes}
\usepackage{slashed}
\usepackage{ textcomp }
\usepackage{float}
\allowdisplaybreaks
\usepackage{xspace}

\definecolor{jd}{rgb}{0.858, 0.188, 0.478}

\def\lapp{\mathrel{\rlap{\raise.5ex\hbox{$<$}}
                    {\lower.5ex\hbox{$\sim$}}}}
\def\gapp{\mathrel{\rlap{\raise.5ex\hbox{$>$}}
                    {\lower.5ex\hbox{$\sim$}}}}

{
{

\newcommand{\bmt}{\begin{pmatrix}}
\newcommand{\emt}{\end{pmatrix}}
\newcommand{\ba}{\begin{array}{c}}
\newcommand{\ea}{\end{array}}
\newcommand{\be}{\begin{equation}}
\newcommand{\ee}{\end{equation}}
\newcommand{\bea}{\begin{eqnarray}}
\newcommand{\eea}{\end{eqnarray}}

\newcommand{\bi}{\begin{itemize}}
\newcommand{\ei}{\end{itemize}}

\newcommand{\baz}{\begin{array}{cc}}
\newcommand{\besub}{\begin{subequations}}
\newcommand{\eesub}{\end{subequations}}

\newcommand{\mathsym}[1]{{}}

\newcommand{\bt}{\begin{tabular}}
\newcommand{\et}{\end{tabular}}

\newcommand{\benu}{\begin{enumerate}}
\newcommand{\eenu}{\end{enumerate}}

\def\a{\alpha}

\def\b{\beta}
\def\g{\gamma}

\def\l{\lambda}
\def\m{\mu}

\def\G{\Gamma}

\def\q2 {q^2}

\def\r {\rightarrow}

\def\bt{\begin{table}}
\def\et{\end{table}}

\newcommand{\bav}{\begin{array}{cccc}}

\title{Freeze-in Dark Matter Through Forbidden Channel in $U(1)_{B-L}$}
\author[a]{Partha Konar,}
\emailAdd{konar@prl.res.in}
\author[a]{Rishav Roshan,}
\emailAdd{rishav@prl.res.in}
\author[a,b]{Sudipta Show}
\emailAdd{sudipta@prl.res.in}
\affiliation[a]{Physical Research Laboratory, Ahmedabad - 380009, Gujarat, India}
\affiliation[b]{Indian Institute of Technology, Gandhinagar - 382424, Gujarat, India}

\abstract{We examine a scenario for freeze-in production of dark matter, which occurs due to the large thermal correction to the mass of a decaying mediator particle present in the thermal bath of the early Universe. We show that the decays, which are kinematically forbidden otherwise, can open up at very high temperatures and dominate the dark matter production. We explore such forbidden production of dark matter in the minimal $U(1)_{B-L}$ model,  comparing dark matter phenomenology in the context of forbidden frozen-in with the standard picture.}

\DeclareUnicodeCharacter{2212}{-}

\begin{document}

\maketitle
\flushbottom

\setcounter{footnote}{0}
\renewcommand*{\thefootnote}{\arabic{footnote}}

\section{Introduction}
The existence of a non-luminous and non-baryonic form of matter popularly known as dark matter (DM) is already well established by the observations like galaxy rotation curve ~\cite{Sofue:2000jx}, Bullet cluster~\cite{Clowe:2006eq}, etc. Besides, the amount of dark matter at present has been measured quite precisely in the experiments like Planck~\cite{Planck:2018vyg}, WMAP~\cite{Hinshaw_2013} by using the Standard Model of cosmology, i.e., the $\Lambda$CDM model. However, the exact nature of dark matter and its production mechanism remain a mystery and an open question. Since the Standard Model(SM) fails to provide a viable DM  candidate, one needs to go beyond the Standard Model(BSM) of particle physics in order to accommodate a feasible dark matter.

Among the existing dark matter models, the most popular ones are those that accommodate dark matter in the form of the weakly interacting massive particle (WIMP)~\cite{Borah:2019aeq, Bhattacharya:2019tqq, Konar:2020wvl, Borah:2020nsz, Konar:2020vuu, DuttaBanik:2020jrj, Chakrabarty:2021kmr, Barman:2021qds, Bhardwaj:2018lma, Bhardwaj:2019mts}. Here, the DM interactions with the SM lies in the weak scale. In the early Universe, such a kind of DM remained in thermal equilibrium with the hot and dense SM plasma. When the temperature of the Universe cooled and fell below its mass, its number density started to dilute, which continued till the time its annihilation rate became smaller than the expansion rate of the Universe. Subsequently, its number density freezes out, and the resulting yield gets saturated.
Interestingly, the dynamics of the freeze-out mechanism can also be affected once the thermal corrections are incorporated, as shown in~\cite{Heurtier:2019beu}. An exciting aspect of the freeze-out is that it is dominated by physics at low energies, as it occurs near a temperature that is 20 - 25 times smaller than the DM mass. Hence the formation mechanism does not carry any information about the history of the early Universe.

Despite such interesting features, the WIMP paradigm still remains in tension due to its null detection at various direct~\cite{Akerib:2016vxi,Zhang:2018xdp,Aprile:2018dbl}, indirect~\cite{MAGIC:2016xys} and collider~\cite{Chatrchyan:2012xdj,Aad:2012tfa} frontiers. These null results motivate us to explore various other possibilities of DM production. Among them, the most popular one is the freeze-in production of the DM from the thermal bath. Here, one assumes: (i) a negligible initial abundance of DM in the early Universe and (ii) a very feeble interaction of the DM with the thermal plasma. Such feeble interactions are responsible for the gradual production of DM from the decay or scattering of the bath particles. Furthermore, the smallness of the same interactions also ensures that the DM never thermalizes with the bath. Once the number density of the particles responsible for the DM production becomes Boltzmann suppressed, the comoving number density of the DM ceases, and its abundance freezes in. This type of DM is popularly known as \emph{feebly interacting massive particle} or FIMP~\cite{Hall:2009bx,Bernal:2017kxu}. Depending on the nature of DM interactions with thermal bath, the freeze-in scenario can be categorized as (i) the Ultra-Violet (UV) freeze-in~\cite{Hall:2009bx,Elahi:2014fsa,Biswas:2019iqm,Barman:2020plp,Barman:2021tgt} and (ii) the Infrared (IR) freeze-in~\cite{Hall:2009bx,Bernal:2017kxu,Biswas:2016bfo,Datta:2021elq,Belanger:2021slj,Bhattacharya:2021jli}.
In a UV freeze-in scenario, the DM interacts with the bath particles only via non-renormalizable ($i.e.$, higher dimensional) operators. In this picture, dark matter is produced solely through the scattering of the bath particles in the early Universe, and the dark matter abundance is highly sensitive to physics at high temperatures, particularly to the reheating temperature of the Universe. Contrary to the UV freeze-in, in IR freeze-in, the dark matter is connected with the visible sector only via renormalizable interaction. Here, dark matter production is dominated primarily at the temperature around the mass of the heavier particle involved in the interactions. 

In this work, we focus on DM production, where it has renormalizable interactions with the thermal bath (IR freeze-in). From here onwards, we will refer to the freeze-in scenario where the DM production dominantly takes place at a temperature near about the mass of the decaying\footnote{Assuming the production of the DM through the scattering of the bath particles remains sub-dominant in comparison the production via decay.} bath particle as a \emph{standard freeze-in} or SFI. Deviating from this scenario, in the present analysis, we consider the production of DM in a parameter regime where its production remains kinematically forbidden in the SFI framework. Darm\'e et~al. studied such a mechanism recently for DM production in Ref.~\cite{Darme:2019wpd}. The interesting feature of this particular production mechanism is the involvement of thermally corrected masses~\cite{Braaten:1993jw, Eberl:2020fml, Rychkov:2007uq, Strumia:2010aa, Darme:2019wpd, Baker:2017zwx, Biondini:2020ric} of the particles participating in the DM production. Here, one considers that the mediator is not only a part of the hot thermal plasma, but it may also acquire a sizable thermal mass. In the early Universe, when the temperature was extremely high, the thermal mass of the mediator can have differed substantially from its mass at vacuum, i.e., the thermal effects must have dominated the mediator's mass. Analogous to the SFI, here, the initial population of dark matter is assumed to be zero or negligibly small, and it is produced gradually from the mediator's decay. At a sufficiently high temperature, the mediator can acquire large thermal mass, and the condition: $M_{\text{mediator}}(T)>2M_{\text{DM}}$ can easily be achieved. The dark matter then can be copiously produced from this decay, even if such a process remains kinematically forbidden at low temperatures. This alternative approach of DM production can be indexed as \emph{forbidden freeze-in} (FFI). This new FFI scenario can open up an exciting and new paradigm of dark matter phenomenology.

This article aims to explore the FFI scenario in a minimal $U(1)_{B-L}$ extension~\cite{PhysRevD.20.776, MARSHAK1980222, Mohapatra:1980qe, PhysRevLett.59.393, Biswas:2016ewm, Biswas:2016bfo, Biswas:2017tce, FileviezPerez:2019cyn, Abdallah:2019svm, Bandyopadhyay:2020ufc} of the SM. As is well known, the $B-L$ extension necessitates the introduction of three right-handed neutrinos (RHN) to make the model free from the triangular anomaly. Unlike the Type-I seesaw~\cite{Minkowski:1977sc, GellMann:1980vs, Mohapatra:1979ia, 10.1143/PTP.64.1103, Schechter:1980gr, Schechter:1981cv, Brdar:2019iem}, here, the bare mass term for the RHNs are not allowed at tree level. Hence, in order to make the RHNs massive, they are required to couple to an SM gauge singlet (complex) scalar appropriately charged under the $U(1)_{B-L}$ symmetry. These RHNs become massive once the $B-L$ scalar acquires a non-zero vacuum expectation value (vev) and spontaneously breaks the $U(1)_{B-L}$ symmetry. In addition, the $B-L$ gauge boson also becomes massive after the breaking of $B-L$ symmetry. It is interesting to point out that the $B-L$ setup can provide a common solution to three of the most important issues of present-day particle physics and cosmology, $, i.e.,$ the non-zero neutrino mass~\cite{Fukuda:1998mi, Ahmad:2002jz, Fogli:2006fw, Ahn:2002up, Brdar:2019iem, Konar:2020vuu}, baryogenesis via leptogenesis~\cite{Fukugita:1986hr, Plumacher:1996kc, Riotto:1999yt, Dine:2003ax, DuttaBanik:2020vfr, Konar:2020vuu, Datta:2021gyi} from the decay of heavier RHNs and dark matter (WIMP/FIMP). The WIMP type DM in the context of the $B-L$ extension has been thoroughly studied ~\cite{Escudero:2018fwn, Basak:2013cga, Bhattacharya:2019tqq}. Here, the lightest RHN (non-trivially charged under a $Z_2$ symmetry) plays the role of a DM~\cite{Okada:2010wd}. Even though such extension can explain all three outstanding issues under the same umbrella, the DM phenomenology still remains highly constrained. The RHN dark matter relic density can only satisfy the Planck limit~\cite{Planck:2018vyg} near the resonance regimes~\cite{Escudero:2018fwn, Basak:2013cga}. An interesting alternative is to consider the lightest RHN as FIMP type DM (SFI). This possibility is also vastly explored in the literature~\cite{Khalil:2008kp, Biswas:2016ewm, Biswas:2016bfo, Biswas:2017tce, Bhattacharya:2019tqq, Seto:2020udg} and unlike the WIMP scenario, here a sizable mass range is allowed\footnote{We would like to point out that in a recent study \cite{Decant:2021mhj}, the authors have shown that Lyman$-\alpha$ bound can also exclude DM mass
		$\lesssim \mathcal{O}(15~\text{keV})$ if produced through a freeze-in mechanism.} for DM. Contrary to this, in the present setup, we follow the FFI approach to study the freeze-in production of dark matter from the kinematically disallowed decay of the scalar that gets a significant thermal mass correction while maintaining equilibrium with the hot thermal plasma in the early Universe.

This paper is organized as follows. In Section~\ref{model}, we introduce the model part while Section~\ref{Thermal_corrections} describes in detail the thermal mass correction of the mediator. Different theoretical and experimental constraints deemed relevant here are described in Section~\ref{constraints}. Next, we present the forbidden freeze-in production of dark matter and the estimation of numerical results in Section~\ref{DMPH} and finally, we summarize our findings in Section~\ref{summary}.
\label{sec:intro}

\section{The scenario}\label{model}
The present scenario explores the possibility of a $U(1)_{B-L}$ extension of the SM gauge symmetry. Here, the particle content is extended by adding three right-handed neutrinos $N_i$ ($i=1,2,3$) together with a complex scalar $S$, all of them charged under the $U(1)_{B-L}$ symmetry. In addition, the SM leptons and quarks also carry $U(1)_{B-L}$ charges of $-1$ and $+\frac{1}{3}$, respectively. Further, invoking an additional unbroken discrete $Z_2$ symmetry and making one of the RHN (say, $N_1$) non-trivially charged under it ensures its stability by forbidding its interactions with the SM leptons and Higgs. Being stable, $N_1$ contributes as a suitable DM candidate in the present setup. On the other hand, the remaining BSM particles and SM particles carry a positive charge under this $Z_2$. In Table~\ref{qno}, we present the charges of all the BSM fields under the different symmetry groups.

\begin{table}[!t]
	\centering
	\begin{tabular}{|c c c c |}
		\hline
		Field   & ~~~~$SU(2)_L \times U(1)_Y$ & ~~~~$Y_{BL}$ & ~~~~$\mathbb{Z}_2$\\ \hline \hline
		$N_1$ & (1,~0) & -1 & $-$         \\ \hline
		$N_2,N_3$ & (1,~0) & -1 & + \\ \hline
		$S$  & (1,~0) & 2 & +             \\ \hline
	\end{tabular}
	\caption{The additional fields and their quantum numbers under different symmetry groups. Here, $Y_{BL}$ refers to the $U(1)_{B-L}$ charge.}
	\label{qno}
\end{table}

These $B-L$ charge assignments also eliminate the possibility of triangular $B-L$ gauge anomalies in our model~\cite{Montero:2007cd}. With the given particle spectrum and the gauge symmetries, the most general renormalizable and gauge invariant Lagrangian for the present setup can be written as,

\bea
\mathcal{L}&=&\mathcal{L}_{KE}+\mathcal{L}_{y}-V(\phi,S)
\label{Lag} 
\eea
where kinetic terms $\mathcal{L}_{KE}$ for the BSM fields are given as,
\bea
\mathcal{L}_{KE} &=& |D_{\mu}S|^2+\sum_{i=1,2,3}\bar{N_i}i\gamma^{\mu}D_{\mu}N_i-\frac{1}{4}Z_{\mu\nu}Z^{\mu\nu}, 
\eea
\text{with}
$Z^{\mu\nu} = \partial^{\mu}Z_{BL}^{\nu}-\partial^{\nu}Z_{BL}^{\mu}$, 
\rm{and} 
$D_{\mu} = \partial_{\mu} + i \, [Yg^{\prime} + Y_{BL} \, g_{BL}] \, (Z_{BL})_{\m}$.
%
Here, we work in the pure $U(1)_{B-L}$ model, where $g^{\prime}$ is considered to be zero. This choice of $g^{\prime}=0$ forbids $Z$-$Z_{BL}$ mixing at the tree level\footnote{The gauge kinetic mixing is highly constrained by electroweak precision measurements demands it to be $\lesssim 10^{-4}$~\cite{Hook:2010tw}.}. Finally, $g_{BL}$ denotes the $U(1)_{B-L}$ gauge coupling.

Moving on to the scalar part of the Lagrangian, the most general renormalizable scalar potential for this setup is given by 

\bea
V(\phi,S) &=& -\mu_{\phi}^2 \phi^{\dagger} \phi - \mu_{S}^2 |S|^2
+ \frac{\l_{\phi}}{2} (\phi^{\dagger} \phi)^2
+ \l_{\phi S} (\phi^{\dagger} \phi) |S|^2 + 
\l_S |S|^4.
\label{potential}
\eea
For $\m_{S}^2>0$, the CP even component of $B-L$ scalar $S= \frac{1}{\sqrt 2}(v_{BL} + \phi_S )$ develops a non-zero vacuum expectation value $v_{BL}$ and breaks the $U(1)_{B-L}$ symmetry. This breaking ensures Majorana masses for the RHNs (discussed latter) together with an additional massive $B-L$ gauge boson $Z_{BL}$. The masses of the $B-L$ scalar  ($\phi_S$) and gauge boson after the $B-L$ symmetry breaking is expressed as~\footnote{After the breaking of $B-L$ symmetry, $\phi$ also obtains mass due to the presence of $\l_{\phi S}$ interaction. We do not write that mass term explicitly as its presence does not alter the present analysis.},
\besub
\bea
m_{S}^2&=& 2 \, \l_S \, v_{BL}^2,\\
M_{Z_{BL}}&=& 2 \, g_{BL} \, v_{BL}.
\eea
\eesub

On the other hand, Electroweak Symmetry Breaking (EWSB) is triggered for $\mu_{\phi}^2$ when the $CP$-even components of $\phi$  receive a vev $v$. The minimization conditions for the potential in Eq.~\ref{potential} are given below:

\besub
\bea
\mu_{\phi}^2 &=& \frac{\l_{\phi}}{2} v^2 + \frac{\l_{\phi S}}{2} v_{BL}^2, \\
\mu_S^2 &=& \frac{\l_{\phi S}}{2} v^2 + \l_S \, v_{BL}^2.
\eea
\eesub
After the EWSB, scalar doublet in the present setup can be parametrized as
\bea
\phi &=&
\begin{pmatrix}
	0\\
	\frac{1}{\sqrt 2}(v + \phi_h)
\end{pmatrix}.
\eea
Subsequent to the EWSB, a non-zero $\phi_h - \phi_S$ mixing leads to the following mass terms
\bea
V \supset \frac{1}{2} \begin{pmatrix}
	\phi_h & \phi_S
\end{pmatrix} \begin{pmatrix}
	\l_{\phi} \, v^2 & \l_{\phi S} \, v \, v_{BL} \\
	\l_{\phi S} \, v \, v_{BL} & 2 \l_S \, v_{BL}^2 
\end{pmatrix} 
\begin{pmatrix}
	\phi_h \\
	\phi_S 
\end{pmatrix}.
\label{scalars}  
\eea
The mass matrix is diagonalised using
\bea
\begin{pmatrix}
	\phi_h \\
	\phi_S
\end{pmatrix} = \begin{pmatrix}
	c_\theta & s_\theta \\
	-s_\theta & c_\theta 
\end{pmatrix}
\begin{pmatrix}
	h \\
	s 
\end{pmatrix}
\eea
with 
\bea
\tan{2\theta} &=& \frac{-2 \, \l_{\phi S} \, v \, v_{BL}}{\l_{\phi} \,  v^2 - 2 \, \l_S \, v^2_{BL} }.
\eea
The mass eigenstates ($h,s$) then have masses
\besub
\bea
m^2_{h,s} &=& \frac{1}{2} \Big[\big(\l_{\phi} v^2 + 2 \l_S v^2_{BL}\big) \pm \sqrt{(\l_\phi v^2 - 2 \l_S v^2_{BL}\big)^2 + 4 \l_{\phi S}^2 v^2 v_{BL}^2}\Big]. 
\eea
\eesub
Here we consider physical scalar $h$ as the SM like Higgs boson with mass $m_h=125.09$ GeV~\cite{deFlorian:2016spz}. 
The various model parameters are expressible in terms of the physical quantities as follows:
\besub
\bea
\l_{\phi} &=& \frac{(m^2_h c^2_\theta + m^2_s s^2_\theta)}{v^2}, \\
\l_{\phi S} &=& \frac{(m^2_s - m^2_h)s_\theta c_\theta}{v \, v_{BL}}, \\
\l_S &=& \frac{(m^2_h s^2_\theta + m^2_s c^2_\theta)}{2 v_{BL}^2}.
\eea
\label{quartic}
\eesub
The $\phi_S-\phi_h$ mixing angle is highly constrained, and the current experiments demand it to be small (see Section~\ref{constraints}). As this mixing angle does not play any significant role in the present context, we have kept $s_{\theta}$ fixed at $10^{-3}$ throughout this work, such that it satisfies the experimental constraints. In the limit of sufficiently small $\phi_S-\phi_h$ mixing, one obtains $\phi_S\simeq s$, and $m_S\simeq m_s$.

Next, the Yukawa interactions for the present scenario is expressed as,
\bea
-\mathcal{L}_{y} \supset 
y_{11} \bar{N_1^c} N_1 S + y_{\a \b} \bar{N_\a^c} N_\b S+h_{i\alpha}\overline{l_L}\tilde{\phi} N_{\alpha}+h.c., 
\label{yuk1}
\eea
with $\a, ~\b = 2,~3$ and $i=e,\m,\tau$. As discussed earlier, $N_1$ being $Z_2$ odd remains stable, unlike the other two RHNs $N_2$ and $N_3$, which can decay into the scalar and the SM leptons ($l$) through the third term of Eq.~\ref{yuk1} if kinematically allowed. The existence of $N_2$ and $N_3$ in the present setup can also explain the origin of non-zero neutrino masses together with baryogenesis via leptogenesis.  
In addition, EWSB gives rise to the following mass matrix for $N_{1,2,3}$.
\bea
M_N = \sqrt{2}~v_{BL}\begin{pmatrix}
	y_{11} & 0 & 0 \\
	0 & y_{22} & y_{23} \\
	0 & y_{23} & y_{33}     
\end{pmatrix}.
\eea
To demonstrate our point without losing the generality, we consider $y_{23} = 0$ for simplicity in the rest of the analysis,
in which case $M_N$ is diagonal with masses 
\bea
M_i =  \sqrt{2}~y_{ii} \, v_{BL}. 
\label{eq:M1}
\eea
For simplicity, we assume the other two RHNs to be nearly mass degenerate for the rest of the analysis and consider $y_{22}\simeq y_{33}=y$. Finally, for our analysis purpose, we choose the following sets of independent parameters: 
\bea
\nonumber
\{m_s,M_1,y,v_{BL}, g_{BL}, s_\theta\}. 
\eea

\section{Thermal corrections}   \label{Thermal_corrections}

This section briefly comments on the thermal corrections to the masses of relevant particles. These corrections play a non-trivial role in understanding the DM phenomenology of the present setup. In the early Universe, when the temperature of the thermal soup was very high, the thermal corrections~\cite{PhysRevD.45.2933, Giudice:2003jh, Baker:2017zwx} to the masses of the particles in the bath must have been very large. In general, any particle that couples in the thermal bath with the primordial plasma is expected to obtain a mass proportional to the temperature of the Universe provided the condition $T>m_i$ is satisfied, here $m_i$ denotes various mass scales involved in the theory~\cite{Comelli:1996vm}.

SM particles are expected to be in equilibrium with the thermal plasma at high temperatures. In the present set up we also assume that the particles like the scalar $S$ and the heavier RHNs $N_{2,3}$ remained in equilibrium with the thermal plasma due to their sizable interaction strengths in the early Universe. Hence, their masses are expected to obtain thermal corrections at high temperature. On the other hand, DM candidate $N_1$ in this model interacts very feebly with the thermal bath and never enters thermal equilibrium. Due to this reason, the thermal correction to its mass remains negligible even at high temperatures. For example, considering $U(1)_{B-L}$ breaking scale $v_{BL}\sim \mathcal{O}(10^{10}~\text{GeV})$ with a fixed DM mass $M_1\sim 500~\text{GeV}$, one obtains $y_{11}\sim 3\times 10^{-8}~\text{GeV}$, following Eq.~\ref{eq:M1}. With such a feeble interaction, the thermal corrections to $N_1$ mass at a temperature $T>>M_1$ remains negligible $i.e.~M_{1}(T)=\sqrt{M_{1}^2+(y_{11}^2/16) T^2}\simeq M_1$. Finally, the setup also demands a very feeble $g_{BL} \sim \mathcal{O}(10^{-8})$. Such a small $g_{BL}$ also prevents $Z_{BL}$ from entering into the equlibrium and hence its thermal mass can also be negelected. These choices of couplings will be further clarified in Section~\ref{DMPH}. 

\begin{figure}[tb!]
\centering
\includegraphics[scale=0.6]{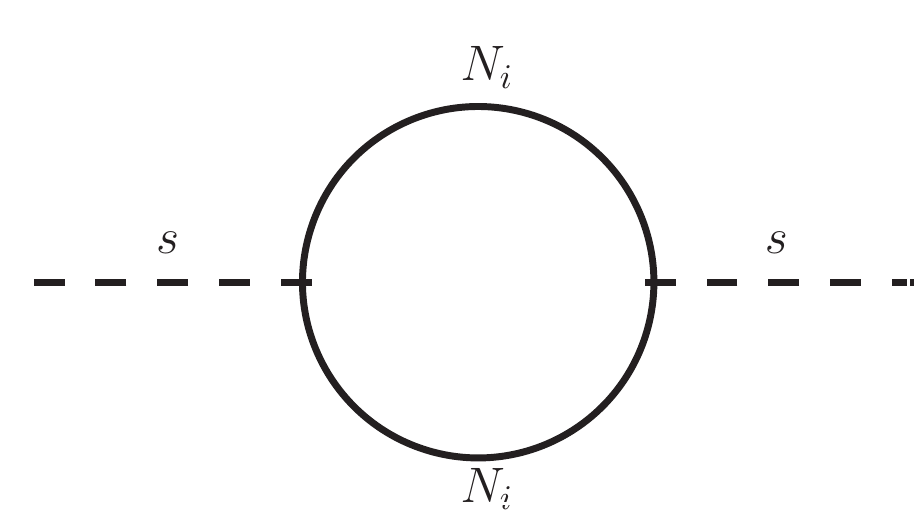}
\caption{One-loop diagram contributing dominantly towards thermally corrected mass of scalar $s$.}
\label{thermal_fig}
\end{figure}

Now we discuss the thermal corrections to the mass of $s$ as it plays a crucial role in the DM phenomenology of the present construct. Note that several processes can provide thermal contributions to the mass of $s$. For example, one can have self-energy corrections with $s$, $H$, $Z_{BL}$ and $N_i$ coming in the loop, which can contribute to the thermally corrected mass of $s$. These contributions can be denoted as $\Pi^2_{s}(T)$, $\Pi^2_{H}(T)$, $\Pi^2_{Z_{Bl}}(T)$ and $\Pi^2_{N_i}(T)$ respectively. The present work demands a very large $v_{BL}$ to ensure the feeble interaction of the DM with $s$. This, in turn, also makes the couplings like $\l_S$ and $\l_{\phi S}$ negligibly small (see Eq.~\ref{quartic}, with $v_{BL}\sim \mathcal{O}(10^{10}~\text{GeV})$, $\l_S\sim 10^{-13}$ and $\l_{\phi S}\sim 10^{-10}$). The smallness of these couplings guarantees that $\Pi^2_{s}$, $\Pi^2_{H}$ remains significantly small in comparison to $\Pi_{N_i}^2(T)$ (Note that the $N_i-N_i-S$ coupling ($y$) can be quite large) and can be ignored. Next, the set up also demands a very small $g_{BL}$ and hence the contributions of $\Pi^2_{Z_{BL}}(T)$ can also be safely ignored. The thermal contribution to the mass of $s$ from the diagram shown in Fig.~\ref{thermal_fig} is given as~\cite{PhysRevD.45.2933, Baker:2017zwx}:
\bea
 \Pi_{N_i}^2(T)&=&\frac{y_{ii}^2}{6}T^2.
\eea
\label{S_corrections}
 Finally, the effective mass of the scalar can be expressed as,

\bea
M_s(T)&=&\sqrt{m_s^2+\Pi_s^2(T)+\Pi_H^2(T)+\Pi_{Z_{BL}}^2(T)+\Pi_{N_i}^2(T)}.
\label{s_thermal_mass}
\eea
%
One can similarly calculate the masses of $N_{2,3}$ in terms of temperature by incorporating all relevant contributions. In Section~\ref{DMPH}, we describe the importance of the thermal corrections in the context of the DM phenomenology.

\section{Theoretical and experimental constraints}\label{constraints}

\subsection{Theoretical constraints}
The scalar potential discussed in Eq.~\ref{potential} must remain bounded from below in various directions in the field space. Stability of vacuum can be ensured if the quartic couplings  satisfy the following conditions:
\bea
\l_{\phi} > 0, 
~ \l_S > 0, 
~ \l_{\phi S} + \sqrt{2\l_{\phi} \l_S} > 0.
\eea
%
On the other hand, to keep the model parameters perturbative, the parameters must
obey:
\bea
|\l_i|< 4\pi,
~ |g_i|< \sqrt{4\pi} ,
~ |y_i|< \sqrt{4\pi},
\eea
Where $g_i$ and $y_i$ denote the gauge, and the Yukawa couplings and $\l_i$ represent the scalar quartic couplings involved in the calculation.

\subsection{Experimental Constraints} 
\begin{itemize}
	
	\item[I.] \textbf{Relic density and direct detection}: Due to the presence of a DM, the model is subjected to the constraints coming from the Planck experiment~\cite{Planck:2018vyg}:
	\bea
	\Omega_{\text{DM}}h^2&=&0.120\pm 0.001.
	\eea 
	Additionally, the model is also exposed to the constraints imposed by the direct detection experiments like  LUX~\cite{Akerib:2016vxi}, PandaX-II~\cite{Zhang:2018xdp} and Xenon-1T~\cite{Aprile:2018dbl}. Elaborated discussions on the dark matter phenomenology are presented in Section~\ref{DMPH}. 
	
	\item[II.]\textbf{LHC diphoton searches}: In presence of the mixing between $h$ and $s$, the tree level interactions of the SM Higgs with the SM fermion and gauge bosons get modified. In such a scenario, the signal strength in the di-photon channel then takes a form:
	\bea
	\mu_{\g \g} = c^2_\theta 
	\frac{BR_{h \to \g \g}}{BR_{h \to \g \g}^{\text{SM}}} \simeq 
	c^2_\theta \frac{\Gamma_{h \to \g \g}}{\Gamma_{h \to \g \g}^{\text{SM}}}.
	\eea
	LHC sets a limit on this new mixing angle as $|\sin{\theta}|\leq 0.36$~\cite{Robens:2016xkb}.
	
	\item[III.]\textbf{LEP bound and opposite sign di-lepton search at LHC}: Since the SM fermions are charged under $U(1)_{B-L}$ symmetry and interact directly with the $U(1)_{B-L}$ gauge boson $Z_{BL}$, the footprints of $Z_{BL}$ can be obtained in the collider searches. The null detection of such signature severely constrains the ratio $M_{Z_{BL}}/g_{BL}$. The exclusion limit from LEP-II~\cite{Carena:2004xs, Cacciapaglia:2006pk} on this ratio is:
	\bea
	\frac{M_{Z_{BL}}}{g_{BL}}\geq 7~\text{TeV}.
	\eea
	On the other hand, one should also observe the constraints coming from opposite-sign di-lepton searches at LHC, which primarily excludes the model for $150~\text{GeV}<M_{Z_{BL}}<3$ TeV~\cite{Escudero:2018fwn, Das:2021esm}, depending on the size of $g_{BL}$. In this work, the $B-L$ gauge boson is treated as a FIMP which in turn demands $g_{BL}$ to be very small, and hence the stringent constraints, as discussed above, can easily be evaded.
	
	\item[IV.]\textbf{Invisible Higgs decay}: In this model, SM Higgs can also decay to the RHNs, $Z_{BL}$ and also to the BSM scalar, if kinematically allowed. These extra decay modes can contribute towards invisible Higgs decay. In such a situation, we need to employ the bound on the invisible Higgs decay width as~\cite{ATLAS:2020kdi}: 
	\besub
	\bea
	Br(h\rightarrow \rm{Invisible})<0.11, \\
	\frac{\Gamma(h\rightarrow \rm{Invisible})}{\Gamma(h\rightarrow SM)+\Gamma(h\rightarrow \rm{Invisible})} < 0.11 .
	\label{invi}   
	\eea 
	\eesub
	where $\Gamma(h\rightarrow \rm{Invisible})=\Gamma(h\rightarrow BSM)$ when $m_{i}<~\frac{m_h}{2}$ with $i=N_1,N_2,N_3,Z_{BL},s$ and $\Gamma(h\rightarrow SM)=4.2$ MeV. However, in our present analysis, we primarily focus on the parameter space where $m_{i}>~\frac{m_h}{2}$. So the above constraint is not applicable.
\end{itemize}

\section{Dark Matter Phenomenology}\label{DMPH}

Null detection of WIMP dark matter in the direct~\cite{Akerib:2016vxi, Zhang:2018xdp, Aprile:2018dbl} and indirect search experiments~\cite{MAGIC:2016xys} has motivated the community to explore the various exotic realization of DM. Among such possibilities, the popular one is the FIMP-type DM, where the DM never comes in equilibrium with the thermal soup. Here, the initial abundance of the DM is assumed to be zero (or negligible). As the Universe cools down, its feeble interaction with the bath helps in its gradual production from decays or scatterings of the bath particles. Such a weaker strength of coupling ensures that the DM interaction rate invariably remains smaller than the Hubble expansion rate ($H$), $i.e.~\G_{int}<H$. Studies of such a FIMP type DM establish a condition where the maximum DM production takes place when the temperature of the thermal bath is of the order or below the mass of the mother particle responsible for the production of the DM. Unlike the standard freeze-in scenario, in the present up, DM production can be enhanced at early times if thermal corrections to the mass of the mother particle are included. This mechanism of DM production can be dubbed as the \emph{forbidden freeze-in}. Here, the DM production channel, which was otherwise forbidden or kinematically disallowed in the standard freeze-in (SFI), now becomes allowed once the thermal correction to the mass of the mother particle is incorporated.

\begin{figure}[tb!]
	\centering
	\includegraphics[scale=0.4]{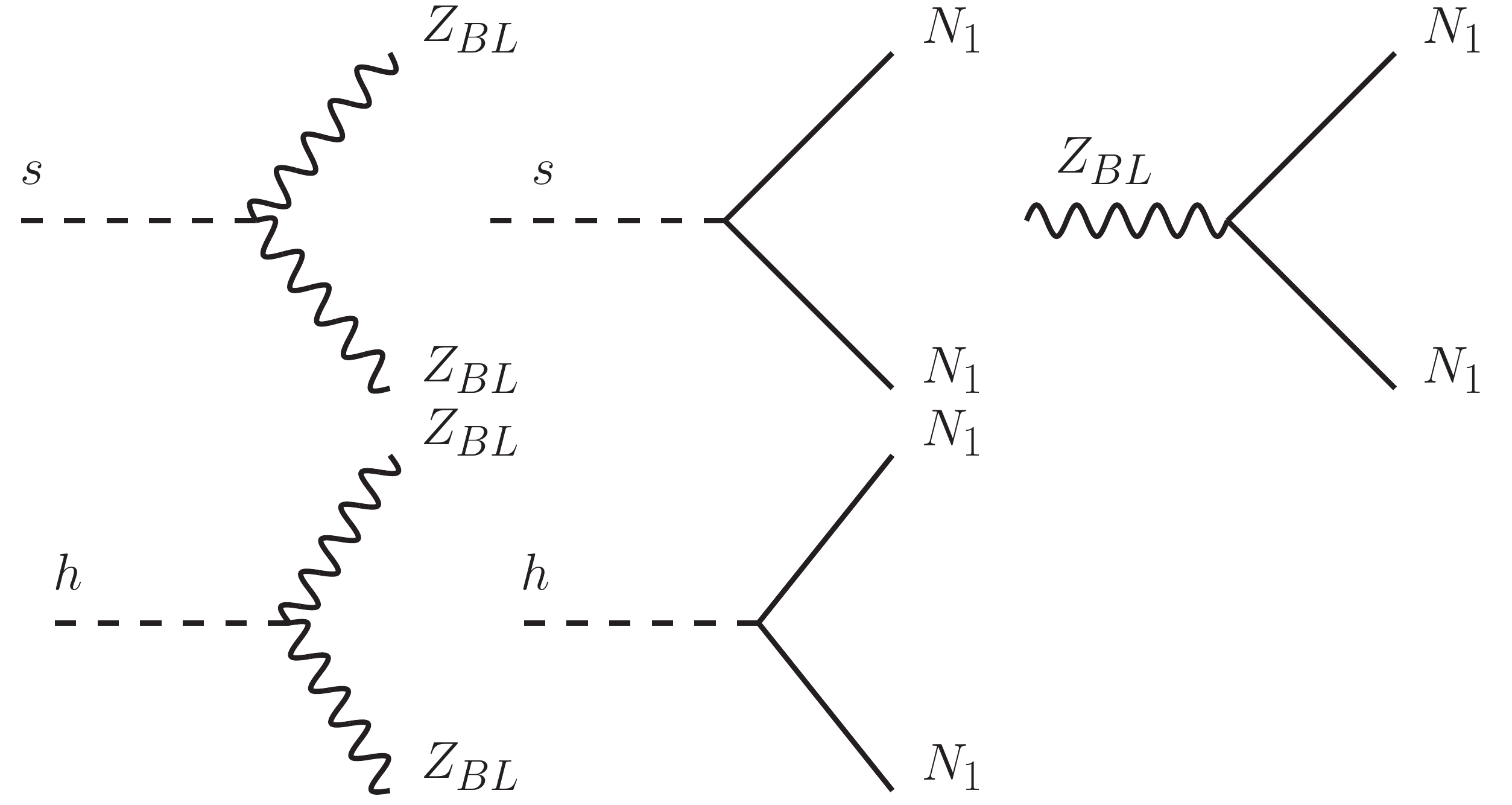}
	\caption{Possible production channels of $Z_{BL}$ and the DM candidate $N_1$.}
	\label{after_EWSB}
\end{figure}

The present setup explores the $U(1)_{B-L}$ extension of the SM where the lightest RHN ($N_1$), which is odd under a $Z_2$ symmetry, plays the role of FIMP dark matter. Here, the production of $N_1$ can take place from the decay of $s$, $h$ (physical scalars obtained after the mixing between $\phi_S$ and $\phi_h$ after the EWSB) and $Z_{BL}$. All such relevant production channels of $Z_{BL}$ and $N_1$ are depicted in Fig.~\ref{after_EWSB}. The feeble interaction of $N_1$ is assured by choosing a relatively large $v_{BL}$ ($\G_{s\to N_1N_1}\propto~y_{11}^2c_{\theta}^2\propto M_1^2c_{\theta}^2/v_{BL}^2$) and a relatively smaller $g_{BL}$ ($\G_{Z_{BL}\to N_1N_1}\propto~g_{BL}^2$). Note that, due to the smallness of $g_{BL}$, the $B-L$ gauge boson $Z_{BL}$ also never thermalizes with bath and is produced feebly from the decay of $s$ and $h$. Hence, in order to study the evolution of dark matter with the expansion of the Universe, one needs to solve a set of coupled Boltzmann equations while taking into account the evolution of $Z_{BL}$ as well. The coupled Boltzmann equations are expressed as,

%
\besub
\bea
\frac{d Y_{Z_{BL}}}{d x} &=& \frac{ 1}{Hx} \Bigg[ \theta(M_s(m_s/x)-2M_{Z_{BL}})~\langle\Gamma_{s \to Z_{BL} Z_{BL}\rangle}Y_{s}^{EQ} 
-\langle\Gamma_{Z_{BL} \to ~all}\rangle Y_{Z_{BL}} \Bigg], \\
\frac{d Y_{N_1}}{d x} &=& \frac{ 1}{Hx}  
\Bigg[
\langle\Gamma_{Z_{BL} \to N_1 N_1}\rangle Y_{Z_{BL}} + \theta(M_s(m_s/x)-2M_{1})~\langle\Gamma_{s \to N_1 N_1}\rangle Y_{s}^{EQ} \Bigg], 
\eea
\label{BE}
\eesub
Here  $x=m_{s}/T$, where $T$ and $H=1.67\sqrt{g_{*}}\frac{T^2}{M_{Pl}}$ denotes the temperature and expansion rate of the Universe respectively. Whereas $Y_j=n_j/\mathfrak{s}$ denotes the comoving number density of the different species ($j=s,Z_{BL},N_1$) involved  with $\mathfrak{s}$ being the entropy density. $Y_s^{EQ}$ signifies the equlibrium density of $s$. Next, $\langle\G_i\rangle$ with $i=s,Z_{BL}$ represents the thermally averaged\footnote{Since $Z_{BL}$ never thermalizes with the plasma, one should properly consider the non-thermal distribution function $(f_{Z_{BL}})$ for $Z_{BL}$ in order to calculate its thermally averaged decay width~\cite{Biswas:2016bfo}. In such a scenario $\langle \Gamma_{Z_{BL}} \rangle=\frac{\int(\frac{M_{Z_{BL}}}{E_{Z_{BL}}})\Gamma_{Z_{BL}\rightarrow AA}f_{Z_{BL}}(p,T)d^3p}{\int f_{Z_{BL}}(p,T)d^3p}$.} decay widths~\cite{Biswas:2016bfo} where 
\besub
\bea
\Gamma_{s \longrightarrow Z_{BL}Z_{BL}} &=& \frac{g_{BL}^2c_{\theta}^2}{8\pi}\frac{M_{s}^3(T)}{M_{Z_{BL}}^2} 
~\Big(1 - \frac{4 M^2_{Z_{BL}}}{M^2_s(T)}\Big)^{1/2}~\Big(1 - \frac{4 M^2_{Z_{BL}}}{M^2_s(T)}+ \frac{12 M^4_{Z_{BL}}}{M^4_s(T)}\Big), \\
\Gamma_{s \longrightarrow N_1 N_1} &=& \frac{M_{s}(T)}{32 \pi} 
~y_{11}^2 c_{\theta}^2 ~\Big(1 - \frac{4 M^2_1}{M^2_s(T)}\Big)^{3/2}, \\
\Gamma_{Z_{BL} \longrightarrow N_1 N_1} &=& \frac{M_{Z_{BL}}}{24 \pi} 
~g^2_{BL} ~\Big(1 - \frac{4 M^2_1}{M^2_{Z_{BL}}}\Big)^{3/2},\\
\Gamma_{Z_{BL} \longrightarrow f \bar{f}} &=& \frac{M_{Z_{BL}}}{12 \pi} 
~g^2_{BL} ~\Big(1 + \frac{2 M^2_f}{M^2_{Z_{BL}}}\Big)
\Big(1 - \frac{4 M^2_f}{M^2_{Z_{BL}}}\Big)^{1/2}. 
\eea
\label{decay_widths}
\eesub
Note that, due to the large $B-L$ breaking scale, BSM particles gain their masses in the early Universe, and hence the $Z_{BL}$ is mainly produced through the decay of $s$. At this stage, we would also like to mention that due to the feeble interaction ($y_{11}$) of the DM with $s$ and large value of both $s$ and $N_i$ masses at high temperature the production of $N_1$ is dominated by the decay of $s$, while its production from scattering processes like $N_iN_i\to N_1N_1$ or $hh (ss)\to N_1N_1$ remains subdominant and can be neglected. Finally, we have also ensured that  rate of the scattering processes like $N_1N_i\to N_1N_i$ and $N_1h(s)\to N_1h(s)$ remains several orders of magnitude smaller than the Hubble expansion rate. For example we found that $ \G_{N_1N_i\to N_1N_i}/H(T)\sim10^{-20}$ at $T\simeq 10^{8}$ GeV which shows that the $N_1$ never enters thermal equilibrium even at high temperatures.

In the present setup, we are interested in exploring the production of both $Z_{BL}$ and $N_1$ through the forbidden channels. These channels become effective once thermal corrections to the mass of $s$ are incorporated and remain active only till the point these decays are kinematically allowed, this is ensured by the use of $\theta-$function in Eq.~\ref{BE}. Once the asymptotic yield of the DM $Y_{N_1}(x_{\infty})$ is obtained after solving the Boltzmann equation, we can use it to calculate the relic density of the DM as,
\bea
\Omega_{N_1}h^2&=&2.75\times 10^8 \bigg(\frac{M_1}{\rm{GeV}}\bigg)Y_{N_1}(x_{\infty}),
\label{relic_expression}
\eea
where $x_{\infty}$ indicates the asymptotic value of $x$ after the DM  freeze-in.

\begin{figure}[tb!]
	\centering
	\includegraphics[scale=0.37]{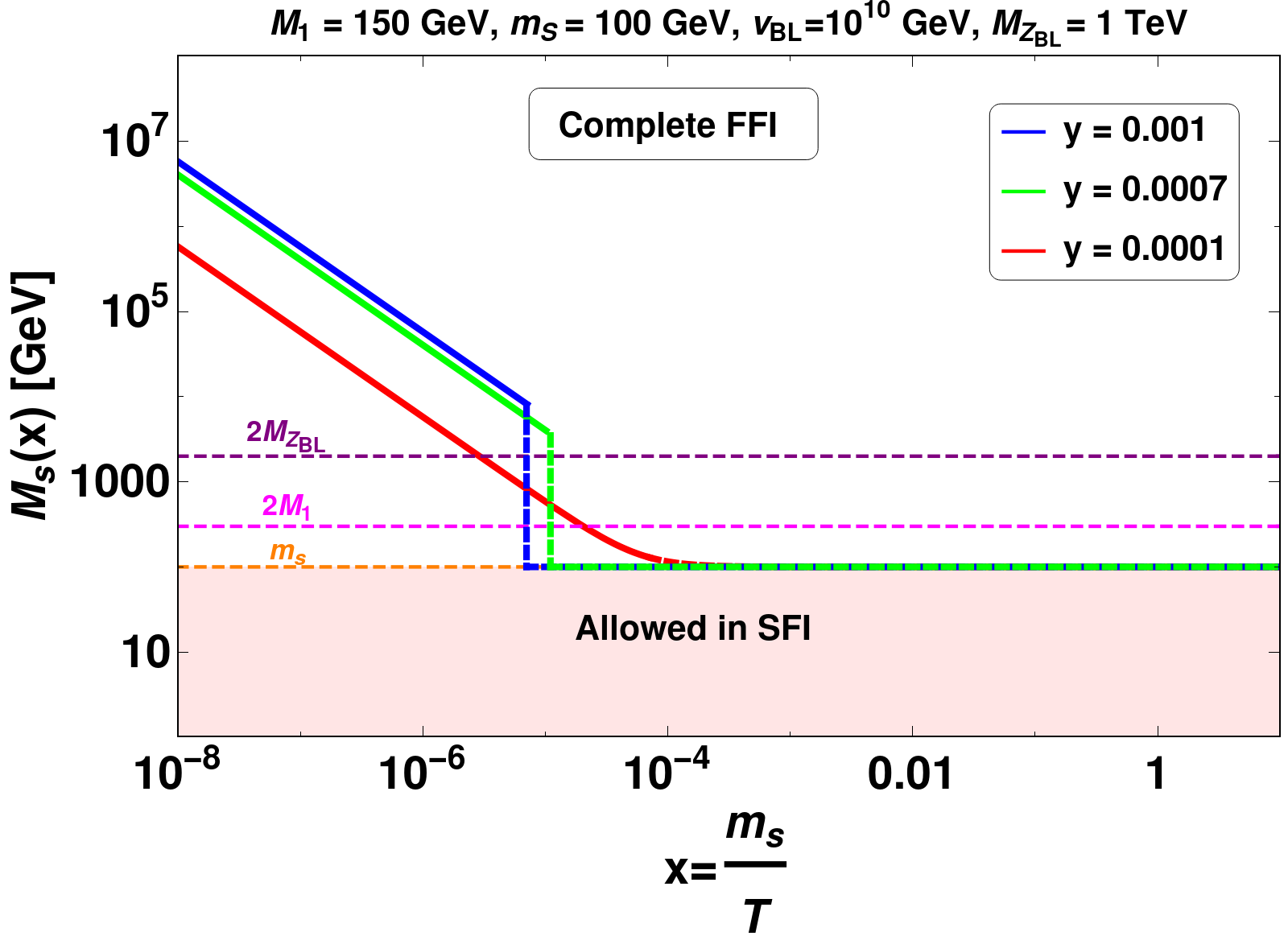}
	\includegraphics[scale=0.37]{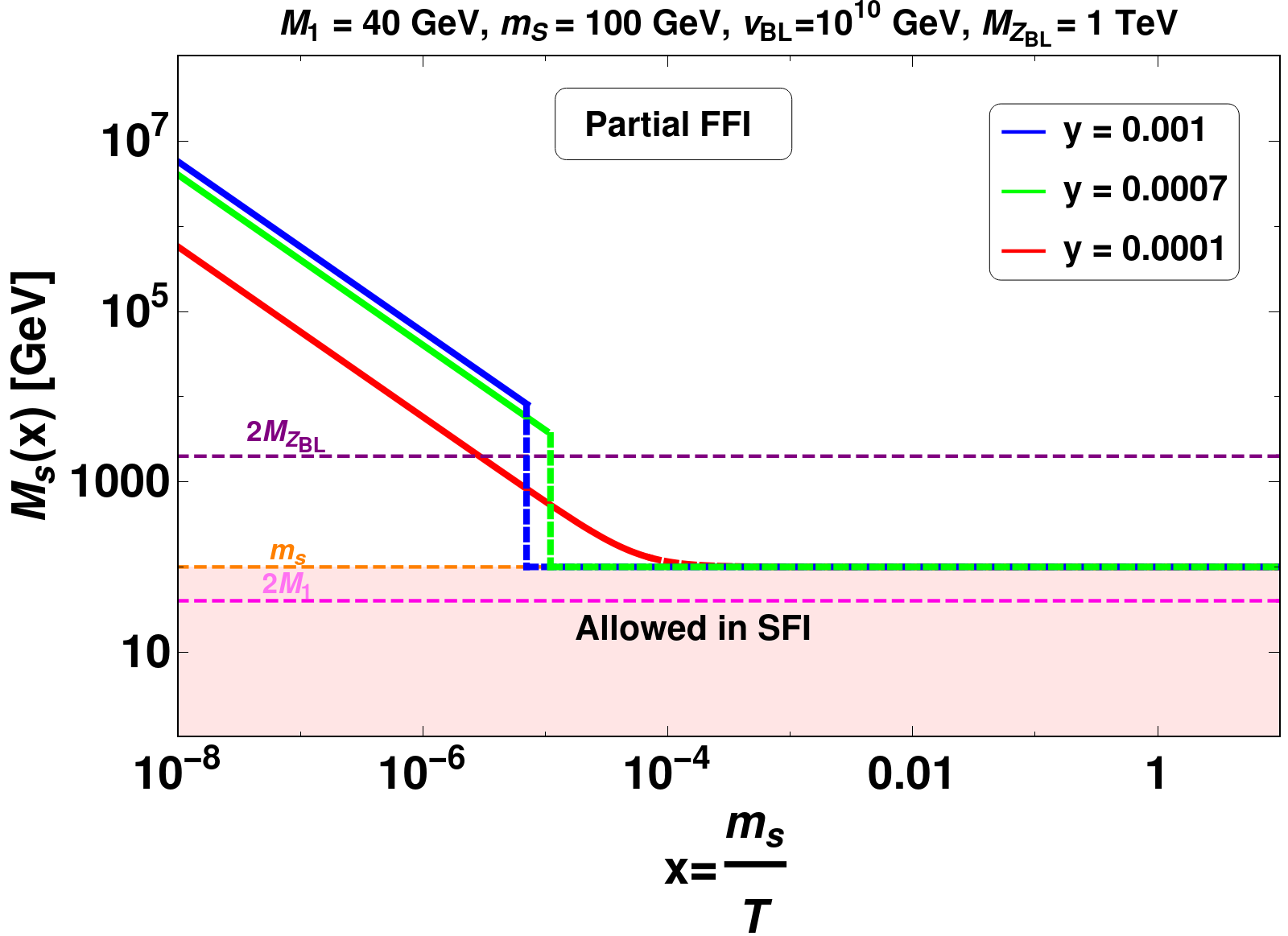}
	\caption{Variation of thermally corrected mass $M_s(x)$ of second scalar with a dimensionless quantity $x=\frac{m_s}{T}$ for three different values Yukawa coupling $y$. In the left panel, we demonstrate a scenario where the $B-L$ gauge boson $Z_{BL}$ and the DM candidate $N_1$ can only be produced via the FFI mechanism. In contrast, the right panel depicts a picture where the $Z_{BL}$ can only be produced via the FFI mechanism, but $N_1$ can be produced through both FFI and SFI.}
	\label{demonstrate}
\end{figure}

	To understand the DM phenomenology more evidently, we categorize our study into two cases in terms of possible mass hierarchies: (A) $M_{Z_{BL}}>M_{1}>m_{s}$, and (B) $M_{Z_{BL}}>m_{s}>M_{1}$ so that the effect of FFI and its benefits over SFI becomes visible. We demonstrate the importance of these two cases in Fig.~\ref{demonstrate}. Here, we show the variation of thermally corrected scalar mass $M_{s}(x)$ in terms of dimensionless parameter $x=m_s/T$ for three different choices of Yukawa couplings $y$.
	Note that, while generating Fig.~\ref{demonstrate} we followed a conservative limit where it is assumed that the thermal correction to the mass of $s$ remains significant till the temperature $T \sim M_{2,3} \sim yv_{BL}$. Below this temperature, the thermally corrected mass of $s$ coincides with the bare mass value~\cite{Comelli:1996vm}. We follow the same principle in presenting the rest of our analysis. 
In Fig.~\ref{demonstrate} the dashed horizontal line represents the fixed values of different mass parameters, $2M_{Z_{BL}}$ (in purple), $2M_{1}$ (in magenta), and $m_s$ (in orange) which helps to understand the mass hierarchy. The pink shaded region shows the parameter space where the FIMP type particles can also be produced if allowed in the SFI scenario. It is evident from the left panel of Fig.~\ref{demonstrate} that the production of the FIMP type particles ($Z_{BL}$ and $N_1$) can only take place through the mechanism of FFI if one considers the mass hierarchy $M_{Z_{BL}}>M_1>m_s$. On contrary to this, in the right panel of Fig.~\ref{demonstrate}, we consider a situation where the mass of the dark matter $i.e.~M_1$ lies below $M_s(T)=m_{s}$. Primary condition on scalar mass $M_{s}(T)=2M_{1}$ ensures the production of the dark matter both from the decay of $s$ and $Z_{BL}$ in the forbidden freeze-in scenario and only through $s$ in a standard freeze-in scenario. 
$s\to Z_{BL}Z_{BL}$ remains forbidden in this case due to the choice of mass hierarchy considered. This case also provides a clear distinction between the FFI and SFI scenarios. Next, we solve the set of coupled Boltzmann equations (Eq.~\ref{BE}) numerically to study the evolution of $Z_{BL}$ and $N_1$ with the expansion of the Universe for these two cases.

\subsection{Case A: Complete FFI region when $M_{Z_{BL}}>M_{1}>m_{s}$ }

In this mass hierarchy, $s$ being the lightest BSM particle, it neither decays to $Z_{BL}$ nor to $N_1$ in a typical SFI scenario. Once the thermally corrected mass of $s$ is taken into account, the left panel of Fig.~\ref{demonstrate} demonstrates that $s$ can be heavy enough to produce both $Z_{BL}$ and $N_1$ through the FFI mechanism. We also provide Table~\ref{tab_casea} for a better understanding of this picture.
\begin{table}[tb!]
	\centering
	\begin{tabular}{|l |c |c | c | c|}
		\hline
		Decay-Channels   & SFI & ~~~~FFI  \\ \hline \hline
		$s\to Z_{BL}Z_{BL}$ & \text{\sffamily X}& \checkmark  \\ \hline
		$s\to N_{1}N_{1}$ & \text{\sffamily X} & \checkmark  \\ \hline
		$Z_{BL}\to N_{1}N_{1}$  & \text{\sffamily X} & \checkmark  \\ \hline
	\end{tabular}
	\caption{List of processes contributing to dark matter and $Z_{BL}$ production in a standard freeze-in (SFI) and forbidden freeze-in (FFI) scenario for a mass hierarchy $M_{Z_{BL}}>M_{1}>m_{s}$. $s \to Z_{BL} Z_{BL}$ remains forbidden within this mass hierarchy for the SFI scenario, which in turn suggests that $Z_{BL}\to N_1N_1$ is also forbidden even though this decay remains kinematically allowed.}
	\label{tab_casea}
\end{table}

\begin{figure}[tb!]
	\centering
	\subfloat[]{\label{caseA_fig_a}\includegraphics[scale=0.35]{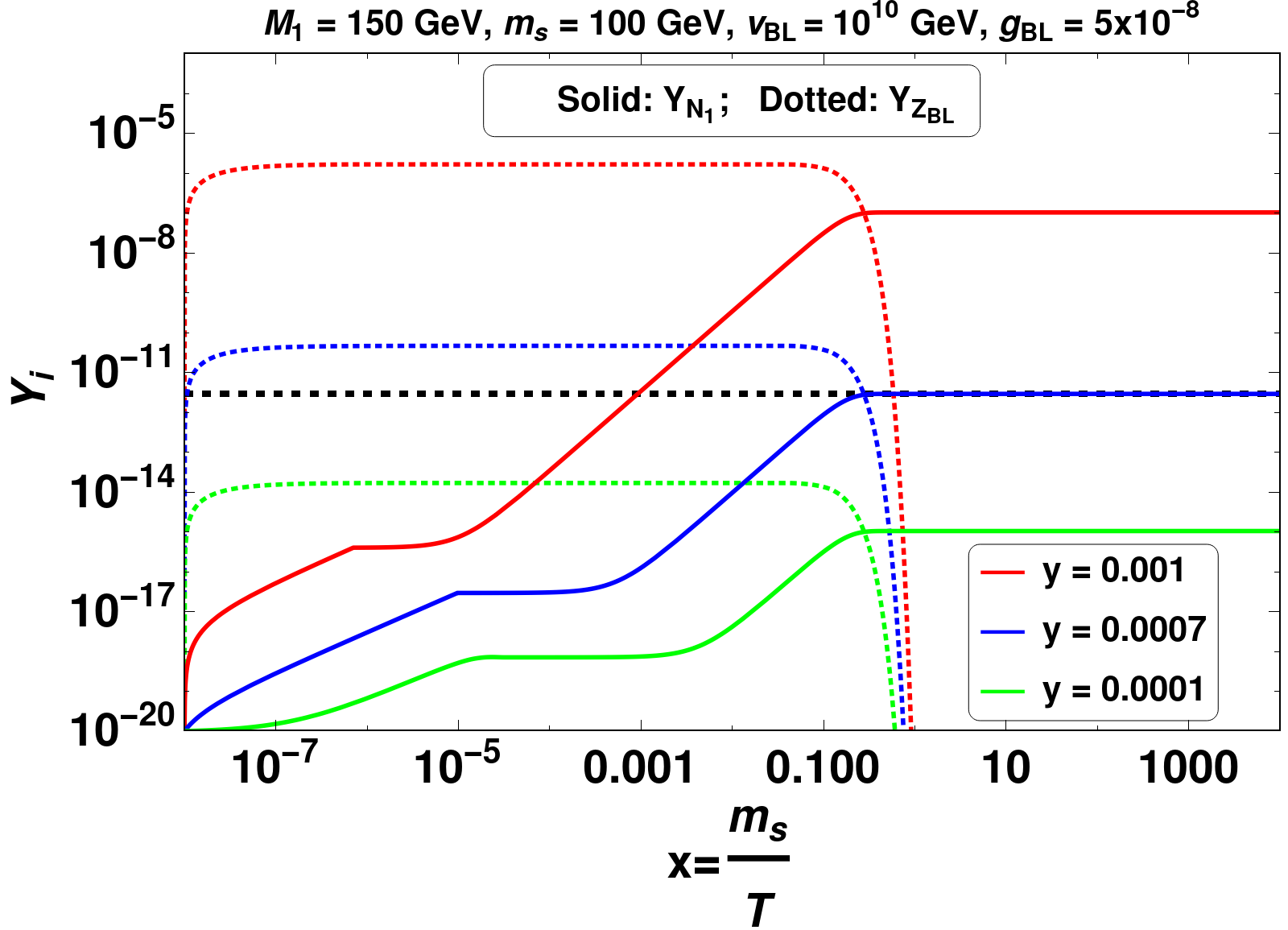}} \hspace{0.4cm}
	\subfloat[]{\label{caseA_fig_b}\includegraphics[scale=0.35]{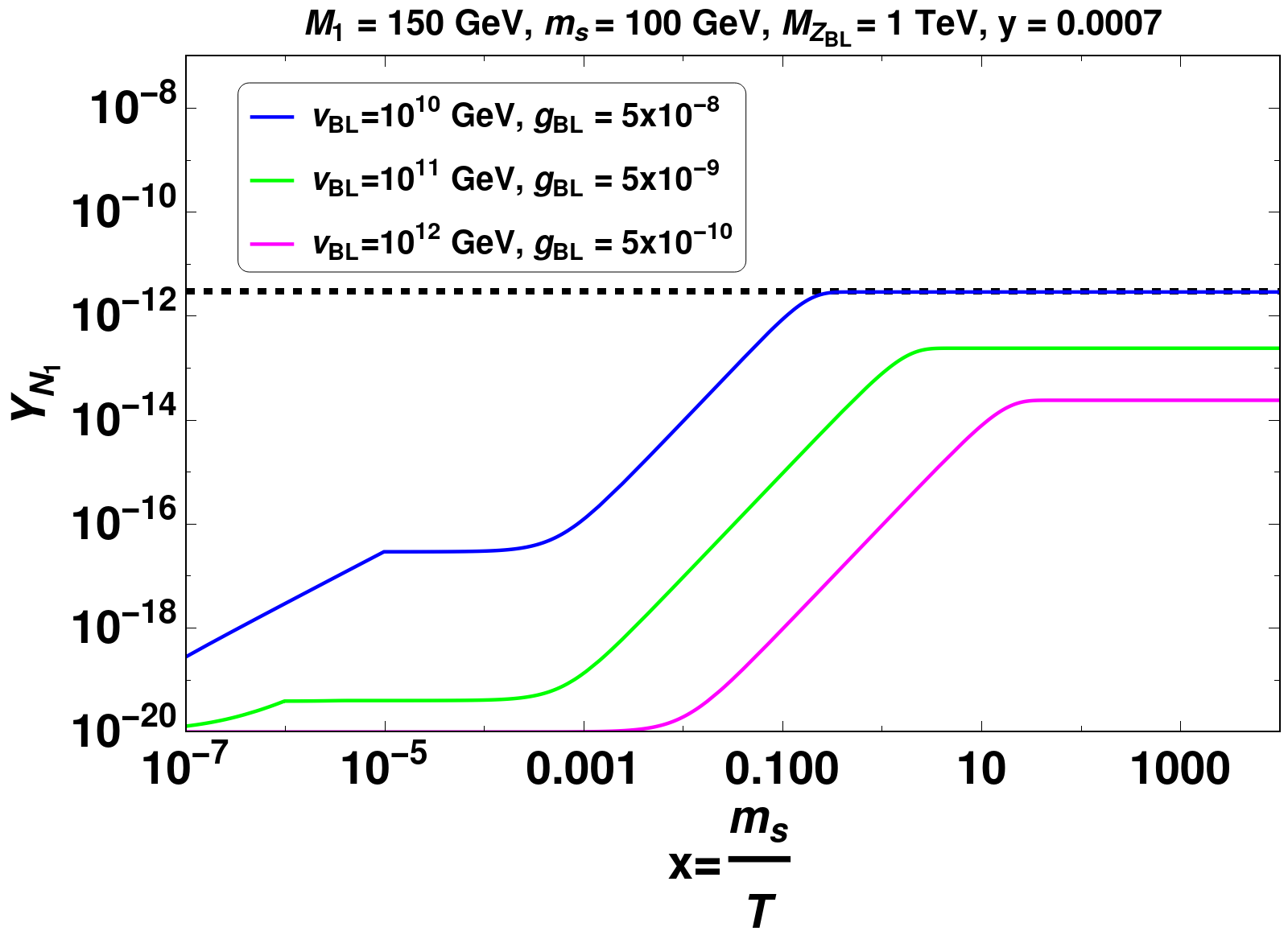}}\\
	\subfloat[]{\label{caseA_fig_c}\includegraphics[scale=0.35]{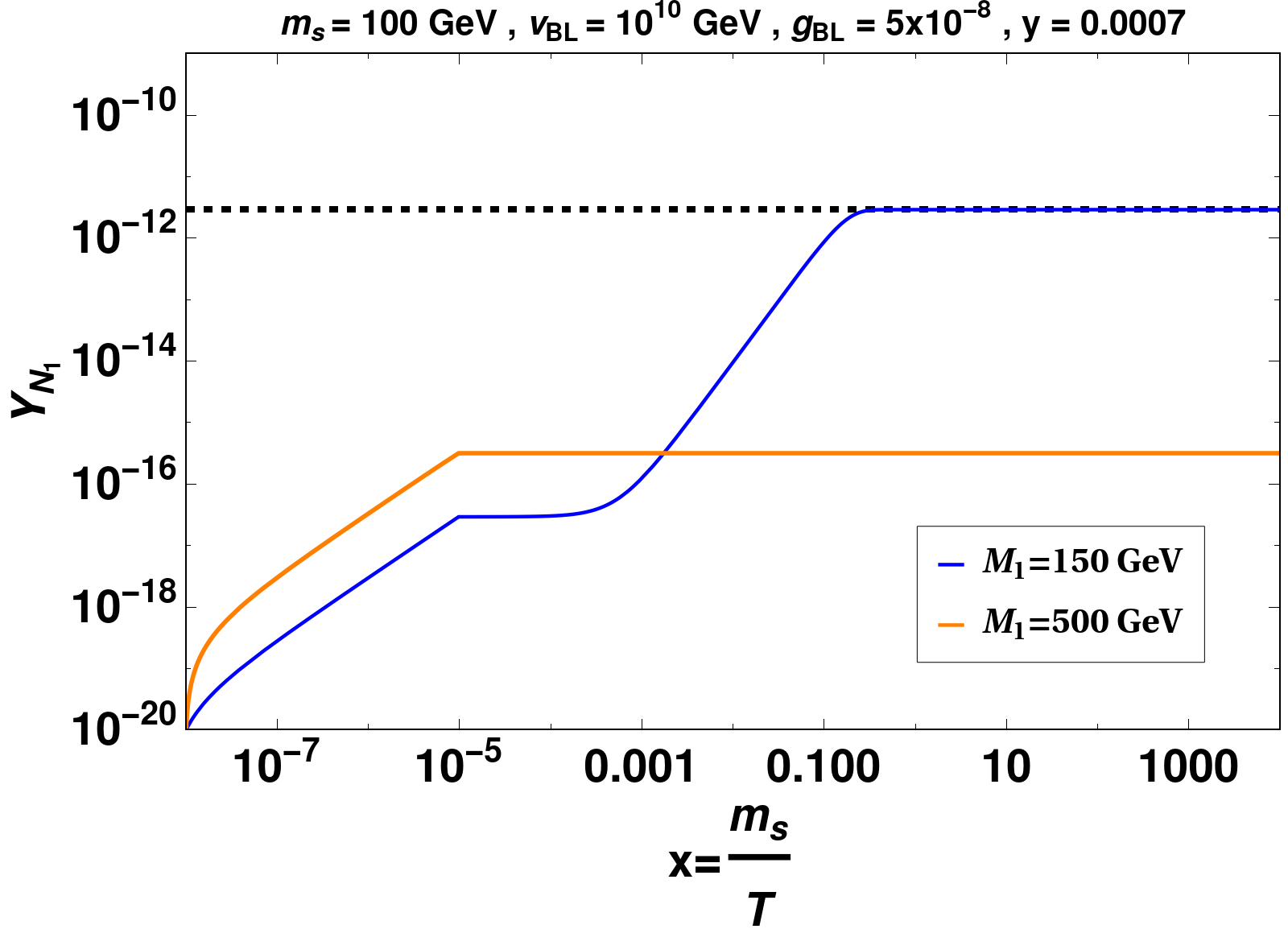}} 
	\caption{Evolution of generated yield of ${Z_{BL}}$ (dotted lines) and dark matter ${N_1}$ (solid lines) with respect to a dimensionless parameter $x=\frac{m_s}{T}$. The values of different parameters controlling the DM phenomenology are mentioned at the top of each plot in a case study for complete FFI region, observed for mass hierarchy $M_{Z_{BL}}>M_{1}>m_{s}$. Thick black dashed line represents the yield of the DM corresponding to the observed relic density. }
	\label{caseA_fig}
\end{figure}

To facilitate our discussion, in Fig.~\ref{caseA_fig_a}, we show the variation of $Y_{Z_{BL}}$ and $Y_{N_1}$ with a dimensionless quantity $x=\frac{m_s}{T}$. The values of different parameters controlling the DM phenomenology are mentioned at the top of each plot. One notices that the production of the $Z_{BL}$ which can occur through the decay of the second scalar $s$ is kinematically forbidden with the given choices of $m_s$ and $M_{Z_{BL}}$ if the thermal corrections are not incorporated. Looking at the Eq.~\ref{quartic}, one finds that for a large $v_{BL}$ as required in this setup, the couplings $\l_{\phi}$ and $\l_{\phi S}$ remains significantly small, on the other hand, the setup also demands a very small $g_{BL}$; hence the contribution of $\Pi_s^2(T),\Pi_H^2(T)$ and $\Pi_{Z_{BL}}^2(T)$ in Eq.~\ref{s_thermal_mass} remains almost negligible. On the other hand, the scalar field can acquire a sizeable thermal mass depending on the choices of BSM Yukawa couplings ($y_{22}~\propto~ M_2/v_{BL}$ and $y_{33}~\propto~ M_3/v_{BL}$). This is also consistent with our expectation that the masses of the other two RHNs must be quite heavy to explain the non-zero neutrino masses and leptogenesis through Type-I seesaw. It is expected that, with an adequate choice of the $y$ (with $y_{22}\simeq y_{33}=y$), one can easily obtain a scenario: $M_{s}(T)>2M_{Z_{BL}},2M_1$ and thereafter, the decay of $s$ can produce $Z_{BL}$ and $N_1$. This can also be seen in Fig.~\ref{caseA_fig_a}. Next, with the choices of parameters considered, the production of $N_{1}$ can also proceed via the decay of $Z_{BL}$. In Fig.~\ref{caseA_fig_a}, the evolution of $Z_{BL}$ (dotted) and $N_1$ (solid) are shown for three different choices of $y$, $i.e.~ y=10^{-3}~(\text{red}),y=7\times10^{-4}~(\text{blue}),y=10^{-4}~(\text{green})$. With $y=10^{-3}$, the thermally corrected mass of $s$ is expected to be large. The larger mass leads to a relatively larger decay widths for the processes $s\to Z_{BL}Z_{BL}$ and $s\to N_1N_1$ (in comparison to smaller $y$ values) which in turn generates relatively larger yields of $Z_{BL}$ and $N_1$ in Fig.~\ref{caseA_fig_a}. One notices that the abundance of $Z_{BL}$ gradually increases due to its production from the decay of $s$, then saturates (plateau) once its production rate becomes comparable to its decay rate. Finally, it falls as its decay to the SM fermions, and the DM overtakes its production. However, the abundance of $N_1$ increases slowly till the time (first bend) when the temperature of the Universe becomes of the order of the heavier RHN masses $i.e.~T\simeq M_i$ ($i.e.~x\simeq 7\times 10^{-6}$ for $y=0.001$), after which its production from the decay of $s$ becomes kinematically forbidden, and its yield saturates. This is because, at this point, the dominant contribution to the thermally corrected mass of $s$ becomes insignificant (as also discussed in Section~\ref{Thermal_corrections}), and $M_s(T)$ falls back to the bare mass value $m_s$. Subsequently, a relatively sharper rise is observed in its yield due to its production from $Z_{BL}$, and finally, its abundance saturates (at around $T \simeq M_{Z_{BL}}~i.e.~x\simeq 10^{-1}$ ) once the decay of $Z_{BL}$ is completed. It is interesting to point out that the production of $Z_{BL}$ from the decay of $s$ starts much earlier in comparison to the production of $N_1$. This happen because the $s$ decays dominantly to $Z_{BL}$ and sub-dominantly to $N_{1}$ (see Eq.~\ref{decay_widths}). Similar behavior is observed in the evolution of $Z_{BL}$ for smaller $y$, but with a relatively smaller yield. With a small $y$, the thermal correction to the mass of $s$ also remains small. This, in turn, reduces the decay width of $s$. Unlike the scenario with a relatively larger $y$, now $N_1$ production ceases when the decay $s \to N_1 N_1$ becomes kinematically disallowed at a relatively later time ( as a smaller $y$ corresponds to a smaller value of $M_{2,3}$, hence a larger $x$). It again starts getting produced as the $Z_{BL}\r N_1N_1$ becomes operational. Finally, DM abundance freezes in once the decay of $Z_{BL}$ is complete. The thick dashed horizontal black line (in each plot) indicates the abundance of dark matter for which the relic density satisfies the Planck experimental limit.

Next, in Fig.~\ref{caseA_fig_b} we show the evolution of the dark matter for three different combinations of $v_{BL}$ and $g_{BL}$ while keeping $M_{Z_{BL}}$ fixed at $1~\text{TeV}$. Here, one finds that for a choice of smaller $v_{BL} $ (and a larger $g_{BL}$), both the FFI production channels get enhanced, leading to an overabundant $N_1$ (as $\G(s \to N_1N_1)\propto\frac{1}{v_{BL}^2}$ and $\G(Z_{BL} \to N_1N_1)\propto g_{BL}^2$). Hence, one can accommodate the correct yield of the DM by tuning these two parameters appropriately, as seen from the blue curve. Lastly, Fig.~\ref{caseA_fig_c} shows the effect of different DM masses on its evolution. For a choice with $M_1=500$ GeV, the only source of its production is the decay of $s$. The moment this decay stops, the DM yield becomes constant. In such a case, it is difficult for the DM to satisfy the measured relic at the Planck experiment.

\subsection{Case B:  Partial FFI region when $M_{Z_{BL}}>m_{s}>M_{1}$ }

\begin{table}[tb!]
	\centering
	\begin{tabular}{|l |c |c | c | c|}
		\hline
		Decay-Channels   & SFI & ~~~~FFI  \\ \hline \hline
		$s\to Z_{BL}Z_{BL}$ & \text{\sffamily X} & \checkmark  \\ \hline
		$s\to N_{1}N_{1}$ & \checkmark & \checkmark  \\ \hline
		$Z_{BL}\to N_{1}N_{1}$  & \text{\sffamily X} & \checkmark  \\ \hline
	\end{tabular}
	\caption{List of processes contributing to dark matter production in a standard freeze-in (SFI) and forbidden freeze-in (FFI) scenario for a mass hierarchy $M_{Z_{BL}}>m_{s}>M_{1}$. $s \to Z_{BL} Z_{BL}$ remains forbidden within this mass hierarchy for the SFI scenario, which in turn suggests that $Z_{BL}\to N_1N_1$ is also forbidden even though this decay remains kinematically allowed.}
	\label{tab_caseb}
\end{table}

\begin{figure}[tb!]
	\centering
	\includegraphics[scale=0.35]{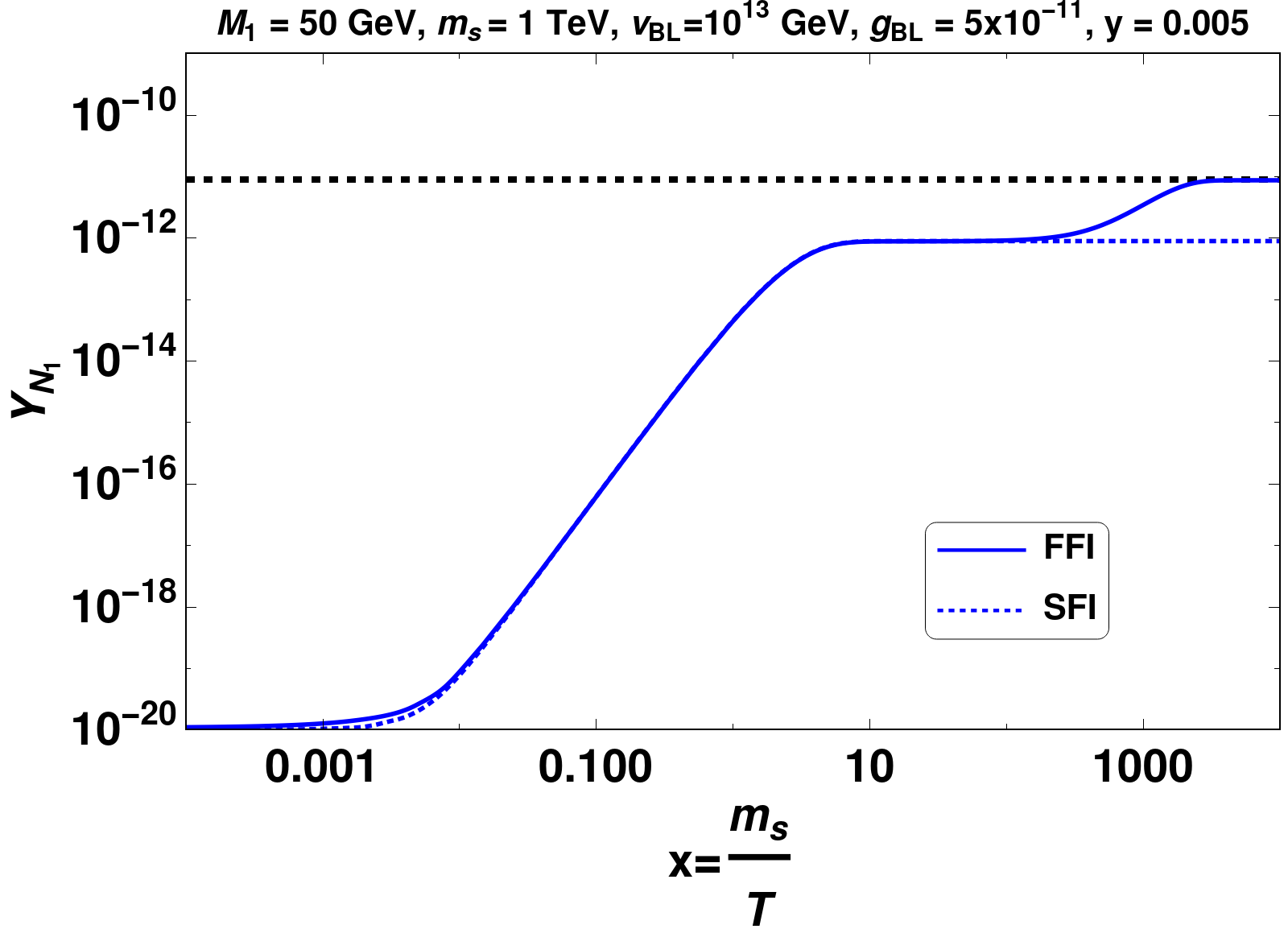}
	\caption{Evolution of generated yield of ${N_1}$ with respect to a dimensionless parameter $x=\frac{m_s}{T}$. The solid blue line depicts the production of $N_1$ in an FFI scenario which can satisfy the Planck experimental limit on the relic density for the given choice of parameters. With the same choice of parameters, the DM remains under-abundant for an SFI scenario, as shown by the dashed blue line. The values of different parameters controlling the DM phenomenology are mentioned at the top of each plot in a case study for partial FFI region, observed for mass hierarchy $M_{Z_{BL}}>m_{s}>M_{1}$. The thick black dashed line represents the abundance of the DM corresponding to the observed relic density. }
	\label{caseB_fig}
\end{figure}

We now aim to study the DM phenomenology with the above mass hierarchy where FFI decay modes are open partially, as also shown in the right panel of Fig~\ref{demonstrate}. Hence evolution process of the DM indicates a distinct direction in the FFI scenario compared to SFI. Unlike the previous case, DM can now be produced directly from the decay of $s$, even if $M_{s}(T)\simeq m_s$ is satisfied. However, the production of the $Z_{BL}$ can only be possible through the forbidden freeze-in mechanism from the decay of $s$\footnote{Although the production of $Z_{BL}$ can proceed through $2\r2$ scatterings, its abundance remains almost negligible as the production cross-section depends on $g_{BL}^4$. }. For a better understanding, in Table~\ref{tab_caseb} we provide all the relevant decay channels required for the production of $Z_{BL}$ and $N_1$ for FFI and SFI. Next, we demonstrate the importance of forbidden freeze-in (FFI) over the standard freeze-in (SSI) in Fig.~\ref{caseB_fig}.


%
This figure shows a comparison between the production of the DM in the SFI scenario (dashed blue line) and the FFI scenario (solid blue line). Here, with the given choice of parameters, the $Z_{BL}$ can never be produced from the decay of $s$ in the SFI scenario. Hence its abundance remains almost negligible (as it can also be produced through scatterings). With such an insignificant yield, $Z_{BL}$ contribution in producing the DM will always remain sub-dominant in comparison to the DM production coming from the $s$ decay. Hence, the DM yield saturates as soon as its production from the $s$ decay stops. In this situation, it may become difficult for the DM to satisfy the correct order of relic density. On the other hand, with the incorporation of FFI, DM can be further produced from the decay of both $s$ and $Z_{BL}$ in an adequate amount to satisfy the correct relic density with the given choice of parameters.

Finally, we also like to comment on the detection prospect of the model under consideration. The spontaneous breaking of the $U(1)_{B-L}$ symmetry at a high energy scale leads to the formation of Nambu-Goto \emph{cosmic strings}~\cite{Bian:2021vmi}. Once formed, the collisions and self-interactions of strings produce non-self-interacting string loops, which further oscillates and radiates their energy in the form of gravitational wave (GW). The incoherent superposition of such continuous emission results in stochastic GW signals. This GW signal can be detected at the present and future GW detectors like pulsar timing arrays (PTAs), NANOGrav~\cite{McLaughlin:2013ira}, PPTA~\cite{Shannon:2015ect}, EPTA~\cite{Kramer:2013kea}, IPTA~\cite{Hobbs:2009yy}, LISA~\cite{LISA:2017pwj}, LIGO~\cite{LIGOScientific:2019vic} etc. The searches of GWs can increase the predictability of the present setup. The detailed study of GWs is beyond the scope of the present work, and we plan to take it as a future endeavor.

\section{Summary and conclusions}  \label{summary}
In this paper, we study the phenomenology of feebly interacting massive particles as dark matter in a minimal $U(1)_{B-L}$ extension of the SM. The role of DM is played by the lightest of the three right-handed neutrinos, which in turn are introduced to make the model free from the triangular anomaly. The other two heavier RHNs can generate non-zero neutrino masses and matter-antimatter asymmetry of the Universe through Type-I seesaw. Here, an unbroken $Z_2$ symmetry ensures the stability of the DM. The setup also requires a complex $SU(2)_L$ singlet scalar charged under the $B-L$ symmetry. After obtaining a non-zero vev, the scalar breaks the $U(1)_{B-L}$ spontaneously and simultaneously makes the RHNs together with a $B-L$ gauge boson massive. 

Due to their feeble interactions with the bath particles, both the DM candidate ($N_1$) and $B-L$ gauge boson ($Z_{BL}$) never comes in equilibrium with the thermal bath. Contrary to this, the complex scalar mediator remains in the thermal equilibrium with the bath due to its not-so-small interactions with the bath particles and contributes to the gradual freeze-in production of $N_1$ and $Z_{BL}$. Moreover, if kinematically allowed, the DM production is further dominated by the $Z_{BL}$ decay.

Although FIMP-type DM is studied in the $B-L$ framework, the thermal corrections to the mediator masses were never taken into account. Incorporating such corrections to the mediator mass at high temperature opens up a new paradigm for a FIMP-type DM phenomenology. Simultaneously, it also opens up an attractive possibility of producing the DM in a kinematically forbidden region of the standard freeze-in (SFI) picture. In this work, we explore this exciting possibility. With this in mind, we categorized our study into two cases depending on the mass hierarchy of these particles. All other likely mass hierarchies can be summed up within these two possibilities.

The first illustration depicts a forbidden freeze-in (FFI) picture where gauge boson and dark matter are heavier than the complex scalar residing in the thermal bath. Hence, the decay is kinematically disallowed, and consequently, such a picture is utterly missing in the SFI framework. The appealing feature here is that the production of the DM can take place in two steps: first from the decay of the scalar due to the thermal corrections to its mass and then subsequently from the late time decay of the gauge boson. Our example explores the synergy between these two processes depending upon parameters in the model.

For further clarity, in our second case study, we choose a particular mass hierarchy ($M_{Z_{BL}}>m_{s}>M_{1}$) to mark the role of FFI over SFI scenarios. Here the production of $Z_{BL}$ is kinematically forbidden in the SFI case, while the dark matter is produced only from $B-L$ scalar's decay. Unlike the standard scenario, with the help of a large thermally corrected mass in FFI, the scalar can produce the gauge boson together with the dark matter in the early Universe. Similar to the first scenario, the DM production again happens in two steps which makes the distinction of FFI with SFI noticeable. Finally, due to the involvement of a large $B-L$ breaking scale, the model can be tested indirectly in the GW search experiments. Such a scale leads to the formation of cosmic string, which further oscillates and radiates its energy in the form of gravitational waves.

\acknowledgments
 This work is supported by the Physical Research Laboratory (PRL), Department of Space, Government of India. Computational work was performed using the HPC resources (Vikram-100 HPC) and TDP project at PRL. Authors gratefully thank H. Mishra for stimulating discussions during the project.

\bibliographystyle{JHEP}
\bibliography{ref}

\providecommand{\href}[2]{#2}\begingroup\raggedright\begin{thebibliography}{10}

\bibitem{Sofue:2000jx}
Y.~Sofue and V.~Rubin, \emph{{Rotation curves of spiral galaxies}},
  \href{https://doi.org/10.1146/annurev.astro.39.1.137}{\emph{Ann. Rev. Astron.
  Astrophys.} {\bfseries 39} (2001) 137}
  [\href{https://arxiv.org/abs/astro-ph/0010594}{{\ttfamily
  astro-ph/0010594}}].

\bibitem{Clowe:2006eq}
D.~Clowe, M.~Bradac, A.~H. Gonzalez, M.~Markevitch, S.~W. Randall, C.~Jones
  et~al., \emph{{A direct empirical proof of the existence of dark matter}},
  \href{https://doi.org/10.1086/508162}{\emph{Astrophys. J. Lett.} {\bfseries
  648} (2006) L109} [\href{https://arxiv.org/abs/astro-ph/0608407}{{\ttfamily
  astro-ph/0608407}}].

\bibitem{Planck:2018vyg}
{\scshape Planck} collaboration, \emph{{Planck 2018 results. VI. Cosmological
  parameters}},
  \href{https://doi.org/10.1051/0004-6361/201833910}{\emph{Astron. Astrophys.}
  {\bfseries 641} (2020) A6}
  [\href{https://arxiv.org/abs/1807.06209}{{\ttfamily 1807.06209}}].

\bibitem{Hinshaw_2013}
G.~Hinshaw, D.~Larson, E.~Komatsu, D.~N. Spergel, C.~L. Bennett, J.~Dunkley
  et~al., \emph{Nine-year wilkinson microwave anisotropy probe ( wmap )
  observations: Cosmological parameter results},
  \href{https://doi.org/10.1088/0067-0049/208/2/19}{\emph{The Astrophysical
  Journal Supplement Series} {\bfseries 208} (2013) 19}.

\bibitem{Borah:2019aeq}
D.~Borah, R.~Roshan and A.~Sil, \emph{{Minimal two-component scalar doublet
  dark matter with radiative neutrino mass}},
  \href{https://doi.org/10.1103/PhysRevD.100.055027}{\emph{Phys. Rev. D}
  {\bfseries 100} (2019) 055027}
  [\href{https://arxiv.org/abs/1904.04837}{{\ttfamily 1904.04837}}].

\bibitem{Bhattacharya:2019tqq}
S.~Bhattacharya, N.~Chakrabarty, R.~Roshan and A.~Sil, \emph{{Multicomponent
  dark matter in extended $U(1)_{B-L}$: neutrino mass and high scale
  validity}}, \href{https://doi.org/10.1088/1475-7516/2020/04/013}{\emph{JCAP}
  {\bfseries 04} (2020) 013}
  [\href{https://arxiv.org/abs/1910.00612}{{\ttfamily 1910.00612}}].

\bibitem{Konar:2020wvl}
P.~Konar, A.~Mukherjee, A.~K. Saha and S.~Show, \emph{{Linking pseudo-Dirac
  dark matter to radiative neutrino masses in a singlet-doublet scenario}},
  \href{https://doi.org/10.1103/PhysRevD.102.015024}{\emph{Phys. Rev. D}
  {\bfseries 102} (2020) 015024}
  [\href{https://arxiv.org/abs/2001.11325}{{\ttfamily 2001.11325}}].

\bibitem{Borah:2020nsz}
D.~Borah, R.~Roshan and A.~Sil, \emph{{Sub-TeV singlet scalar dark matter and
  electroweak vacuum stability with vectorlike fermions}},
  \href{https://doi.org/10.1103/PhysRevD.102.075034}{\emph{Phys. Rev. D}
  {\bfseries 102} (2020) 075034}
  [\href{https://arxiv.org/abs/2007.14904}{{\ttfamily 2007.14904}}].

\bibitem{Konar:2020vuu}
P.~Konar, A.~Mukherjee, A.~K. Saha and S.~Show, \emph{{A dark clue to seesaw
  and leptogenesis in a pseudo-Dirac singlet doublet scenario with
  (non)standard cosmology}},
  \href{https://doi.org/10.1007/JHEP03(2021)044}{\emph{JHEP} {\bfseries 03}
  (2021) 044} [\href{https://arxiv.org/abs/2007.15608}{{\ttfamily
  2007.15608}}].

\bibitem{DuttaBanik:2020jrj}
A.~Dutta~Banik, R.~Roshan and A.~Sil, \emph{{Two component singlet-triplet
  scalar dark matter and electroweak vacuum stability}},
  \href{https://doi.org/10.1103/PhysRevD.103.075001}{\emph{Phys. Rev. D}
  {\bfseries 103} (2021) 075001}
  [\href{https://arxiv.org/abs/2009.01262}{{\ttfamily 2009.01262}}].

\bibitem{Chakrabarty:2021kmr}
N.~Chakrabarty, R.~Roshan and A.~Sil, \emph{{Two Component Doublet-Triplet
  Scalar Dark Matter stabilising the Electroweak vacuum}},
  \href{https://arxiv.org/abs/2102.06032}{{\ttfamily 2102.06032}}.

\bibitem{Barman:2021qds}
B.~Barman, N.~Bernal, A.~Das and R.~Roshan, \emph{{Non-minimally Coupled Vector
  Boson Dark Matter}},  \href{https://arxiv.org/abs/2108.13447}{{\ttfamily
  2108.13447}}.

\bibitem{Bhardwaj:2018lma}
A.~Bhardwaj, A.~Das, P.~Konar and A.~Thalapillil, \emph{{Looking for Minimal
  Inverse Seesaw scenarios at the LHC with Jet Substructure Techniques}},
  \href{https://doi.org/10.1088/1361-6471/ab7769}{\emph{J. Phys. G} {\bfseries
  47} (2020) 075002} [\href{https://arxiv.org/abs/1801.00797}{{\ttfamily
  1801.00797}}].

\bibitem{Bhardwaj:2019mts}
A.~Bhardwaj, P.~Konar, T.~Mandal and S.~Sadhukhan, \emph{{Probing the inert
  doublet model using jet substructure with a multivariate analysis}},
  \href{https://doi.org/10.1103/PhysRevD.100.055040}{\emph{Phys. Rev. D}
  {\bfseries 100} (2019) 055040}
  [\href{https://arxiv.org/abs/1905.04195}{{\ttfamily 1905.04195}}].

\bibitem{Heurtier:2019beu}
L.~Heurtier and H.~Partouche, \emph{{Spontaneous Freeze Out of Dark Matter From
  an Early Thermal Phase Transition}},
  \href{https://doi.org/10.1103/PhysRevD.101.043527}{\emph{Phys. Rev. D}
  {\bfseries 101} (2020) 043527}
  [\href{https://arxiv.org/abs/1912.02828}{{\ttfamily 1912.02828}}].

\bibitem{Akerib:2016vxi}
{\scshape LUX} collaboration, \emph{{Results from a search for dark matter in
  the complete LUX exposure}},
  \href{https://doi.org/10.1103/PhysRevLett.118.021303}{\emph{Phys. Rev. Lett.}
  {\bfseries 118} (2017) 021303}
  [\href{https://arxiv.org/abs/1608.07648}{{\ttfamily 1608.07648}}].

\bibitem{Zhang:2018xdp}
{\scshape PandaX} collaboration, \emph{{Dark matter direct search sensitivity
  of the PandaX-4T experiment}},
  \href{https://doi.org/10.1007/s11433-018-9259-0}{\emph{Sci. China Phys. Mech.
  Astron.} {\bfseries 62} (2019) 31011}
  [\href{https://arxiv.org/abs/1806.02229}{{\ttfamily 1806.02229}}].

\bibitem{Aprile:2018dbl}
{\scshape XENON} collaboration, \emph{{Dark Matter Search Results from a One
  Ton-Year Exposure of XENON1T}},
  \href{https://doi.org/10.1103/PhysRevLett.121.111302}{\emph{Phys. Rev. Lett.}
  {\bfseries 121} (2018) 111302}
  [\href{https://arxiv.org/abs/1805.12562}{{\ttfamily 1805.12562}}].

\bibitem{MAGIC:2016xys}
{\scshape MAGIC, Fermi-LAT} collaboration, \emph{{Limits to Dark Matter
  Annihilation Cross-Section from a Combined Analysis of MAGIC and Fermi-LAT
  Observations of Dwarf Satellite Galaxies}},
  \href{https://doi.org/10.1088/1475-7516/2016/02/039}{\emph{JCAP} {\bfseries
  02} (2016) 039} [\href{https://arxiv.org/abs/1601.06590}{{\ttfamily
  1601.06590}}].

\bibitem{Chatrchyan:2012xdj}
{\scshape CMS} collaboration, \emph{{Observation of a New Boson at a Mass of
  125 GeV with the CMS Experiment at the LHC}},
  \href{https://doi.org/10.1016/j.physletb.2012.08.021}{\emph{Phys. Lett.}
  {\bfseries B716} (2012) 30}
  [\href{https://arxiv.org/abs/1207.7235}{{\ttfamily 1207.7235}}].

\bibitem{Aad:2012tfa}
{\scshape ATLAS} collaboration, \emph{{Observation of a new particle in the
  search for the Standard Model Higgs boson with the ATLAS detector at the
  LHC}}, \href{https://doi.org/10.1016/j.physletb.2012.08.020}{\emph{Phys.
  Lett.} {\bfseries B716} (2012) 1}
  [\href{https://arxiv.org/abs/1207.7214}{{\ttfamily 1207.7214}}].

\bibitem{Hall:2009bx}
L.~J. Hall, K.~Jedamzik, J.~March-Russell and S.~M. West, \emph{{Freeze-In
  Production of FIMP Dark Matter}},
  \href{https://doi.org/10.1007/JHEP03(2010)080}{\emph{JHEP} {\bfseries 03}
  (2010) 080} [\href{https://arxiv.org/abs/0911.1120}{{\ttfamily 0911.1120}}].

\bibitem{Bernal:2017kxu}
N.~Bernal, M.~Heikinheimo, T.~Tenkanen, K.~Tuominen and V.~Vaskonen, \emph{{The
  Dawn of FIMP Dark Matter: A Review of Models and Constraints}},
  \href{https://doi.org/10.1142/S0217751X1730023X}{\emph{Int. J. Mod. Phys. A}
  {\bfseries 32} (2017) 1730023}
  [\href{https://arxiv.org/abs/1706.07442}{{\ttfamily 1706.07442}}].

\bibitem{Elahi:2014fsa}
F.~Elahi, C.~Kolda and J.~Unwin, \emph{{UltraViolet Freeze-in}},
  \href{https://doi.org/10.1007/JHEP03(2015)048}{\emph{JHEP} {\bfseries 03}
  (2015) 048} [\href{https://arxiv.org/abs/1410.6157}{{\ttfamily 1410.6157}}].

\bibitem{Biswas:2019iqm}
A.~Biswas, S.~Ganguly and S.~Roy, \emph{{Fermionic dark matter via UV and IR
  freeze-in and its possible X-ray signature}},
  \href{https://doi.org/10.1088/1475-7516/2020/03/043}{\emph{JCAP} {\bfseries
  03} (2020) 043} [\href{https://arxiv.org/abs/1907.07973}{{\ttfamily
  1907.07973}}].

\bibitem{Barman:2020plp}
B.~Barman, D.~Borah and R.~Roshan, \emph{{Effective Theory of Freeze-in Dark
  Matter}}, \href{https://doi.org/10.1088/1475-7516/2020/11/021}{\emph{JCAP}
  {\bfseries 11} (2020) 021}
  [\href{https://arxiv.org/abs/2007.08768}{{\ttfamily 2007.08768}}].

\bibitem{Barman:2021tgt}
B.~Barman, D.~Borah and R.~Roshan, \emph{{Nonthermal leptogenesis and UV
  freeze-in of dark matter: Impact of inflationary reheating}},
  \href{https://doi.org/10.1103/PhysRevD.104.035022}{\emph{Phys. Rev. D}
  {\bfseries 104} (2021) 035022}
  [\href{https://arxiv.org/abs/2103.01675}{{\ttfamily 2103.01675}}].

\bibitem{Biswas:2016bfo}
A.~Biswas and A.~Gupta, \emph{{Freeze-in Production of Sterile Neutrino Dark
  Matter in U(1)$_{\rm B-L}$ Model}},
  \href{https://doi.org/10.1088/1475-7516/2016/09/044}{\emph{JCAP} {\bfseries
  09} (2016) 044} [\href{https://arxiv.org/abs/1607.01469}{{\ttfamily
  1607.01469}}].

\bibitem{Datta:2021elq}
A.~Datta, R.~Roshan and A.~Sil, \emph{{Imprint of the seesaw mechanism on
  feebly interacting dark matter and the baryon asymmetry}},
  \href{https://arxiv.org/abs/2104.02030}{{\ttfamily 2104.02030}}.

\bibitem{Belanger:2021slj}
G.~B\'elanger, S.~Khan, R.~Padhan, M.~Mitra and S.~Shil, \emph{{Right handed
  neutrinos, TeV scale BSM neutral Higgs boson, and FIMP dark matter in an EFT
  framework}}, \href{https://doi.org/10.1103/PhysRevD.104.055047}{\emph{Phys.
  Rev. D} {\bfseries 104} (2021) 055047}
  [\href{https://arxiv.org/abs/2104.04373}{{\ttfamily 2104.04373}}].

\bibitem{Bhattacharya:2021jli}
S.~Bhattacharya, R.~Roshan, A.~Sil and D.~Vatsyayan, \emph{{Symmetry origin of
  Baryon Asymmetry, Dark Matter and Neutrino Mass}},
  \href{https://arxiv.org/abs/2105.06189}{{\ttfamily 2105.06189}}.

\bibitem{Darme:2019wpd}
L.~Darm\'e, A.~Hryczuk, D.~Karamitros and L.~Roszkowski, \emph{{Forbidden
  frozen-in dark matter}},
  \href{https://doi.org/10.1007/JHEP11(2019)159}{\emph{JHEP} {\bfseries 11}
  (2019) 159} [\href{https://arxiv.org/abs/1908.05685}{{\ttfamily
  1908.05685}}].

\bibitem{Braaten:1993jw}
E.~Braaten and D.~Segel, \emph{{Neutrino energy loss from the plasma process at
  all temperatures and densities}},
  \href{https://doi.org/10.1103/PhysRevD.48.1478}{\emph{Phys. Rev. D}
  {\bfseries 48} (1993) 1478}
  [\href{https://arxiv.org/abs/hep-ph/9302213}{{\ttfamily hep-ph/9302213}}].

\bibitem{Eberl:2020fml}
H.~Eberl, I.~D. Gialamas and V.~C. Spanos, \emph{{Gravitino thermal production
  revisited}}, \href{https://doi.org/10.1103/PhysRevD.103.075025}{\emph{Phys.
  Rev. D} {\bfseries 103} (2021) 075025}
  [\href{https://arxiv.org/abs/2010.14621}{{\ttfamily 2010.14621}}].

\bibitem{Rychkov:2007uq}
V.~S. Rychkov and A.~Strumia, \emph{{Thermal production of gravitinos}},
  \href{https://doi.org/10.1103/PhysRevD.75.075011}{\emph{Phys. Rev. D}
  {\bfseries 75} (2007) 075011}
  [\href{https://arxiv.org/abs/hep-ph/0701104}{{\ttfamily hep-ph/0701104}}].

\bibitem{Strumia:2010aa}
A.~Strumia, \emph{{Thermal production of axino Dark Matter}},
  \href{https://doi.org/10.1007/JHEP06(2010)036}{\emph{JHEP} {\bfseries 06}
  (2010) 036} [\href{https://arxiv.org/abs/1003.5847}{{\ttfamily 1003.5847}}].

\bibitem{Baker:2017zwx}
M.~J. Baker, M.~Breitbach, J.~Kopp and L.~Mittnacht, \emph{{Dynamic Freeze-In:
  Impact of Thermal Masses and Cosmological Phase Transitions on Dark Matter
  Production}}, \href{https://doi.org/10.1007/JHEP03(2018)114}{\emph{JHEP}
  {\bfseries 03} (2018) 114}
  [\href{https://arxiv.org/abs/1712.03962}{{\ttfamily 1712.03962}}].

\bibitem{Biondini:2020ric}
S.~Biondini and J.~Ghiglieri, \emph{{Freeze-in produced dark matter in the
  ultra-relativistic regime}},
  \href{https://doi.org/10.1088/1475-7516/2021/03/075}{\emph{JCAP} {\bfseries
  03} (2021) 075} [\href{https://arxiv.org/abs/2012.09083}{{\ttfamily
  2012.09083}}].

\bibitem{PhysRevD.20.776}
A.~Davidson, \emph{$b\ensuremath{-}l$ as the fourth color within an
  $\mathrm{SU}{(2)}_{L}\ifmmode\times\else\texttimes\fi{}\mathrm{U}{(1)}_{R}\ifmmode\times\else\texttimes\fi{}\mathrm{U}(1)$
  model}, \href{https://doi.org/10.1103/PhysRevD.20.776}{\emph{Phys. Rev. D}
  {\bfseries 20} (1979) 776}.

\bibitem{MARSHAK1980222}
R.~Marshak and R.~Mohapatra, \emph{Quark-lepton symmetry and b − l as the
  u(1) generator of the electroweak symmetry group},
  \href{https://doi.org/https://doi.org/10.1016/0370-2693(80)90436-0}{\emph{Physics
  Letters B} {\bfseries 91} (1980) 222}.

\bibitem{Mohapatra:1980qe}
R.~N. Mohapatra and R.~E. Marshak, \emph{{Local B-L Symmetry of Electroweak
  Interactions, Majorana Neutrinos and Neutron Oscillations}},
  \href{https://doi.org/10.1103/PhysRevLett.44.1316}{\emph{Phys. Rev. Lett.}
  {\bfseries 44} (1980) 1316}.

\bibitem{PhysRevLett.59.393}
A.~Davidson and K.~C. Wali, \emph{Universal seesaw mechanism?},
  \href{https://doi.org/10.1103/PhysRevLett.59.393}{\emph{Phys. Rev. Lett.}
  {\bfseries 59} (1987) 393}.

\bibitem{Biswas:2016ewm}
A.~Biswas, S.~Choubey and S.~Khan, \emph{{Galactic gamma ray excess and dark
  matter phenomenology in a $U(1)_{B-L}$ model}},
  \href{https://doi.org/10.1007/JHEP08(2016)114}{\emph{JHEP} {\bfseries 08}
  (2016) 114} [\href{https://arxiv.org/abs/1604.06566}{{\ttfamily
  1604.06566}}].

\bibitem{Biswas:2017tce}
A.~Biswas, S.~Choubey and S.~Khan, \emph{{Neutrino mass, leptogenesis and FIMP
  dark matter in a $\mathrm{U}(1)_{B-L}$ model}},
  \href{https://doi.org/10.1140/epjc/s10052-017-5436-y}{\emph{Eur. Phys. J. C}
  {\bfseries 77} (2017) 875}
  [\href{https://arxiv.org/abs/1704.00819}{{\ttfamily 1704.00819}}].

\bibitem{FileviezPerez:2019cyn}
P.~Fileviez~P\'erez, C.~Murgui and A.~D. Plascencia, \emph{{Neutrino-Dark
  Matter Connections in Gauge Theories}},
  \href{https://doi.org/10.1103/PhysRevD.100.035041}{\emph{Phys. Rev. D}
  {\bfseries 100} (2019) 035041}
  [\href{https://arxiv.org/abs/1905.06344}{{\ttfamily 1905.06344}}].

\bibitem{Abdallah:2019svm}
W.~Abdallah, S.~Choubey and S.~Khan, \emph{{FIMP dark matter candidate(s) in a
  $B − L$ model with inverse seesaw mechanism}},
  \href{https://doi.org/10.1007/JHEP06(2019)095}{\emph{JHEP} {\bfseries 06}
  (2019) 095} [\href{https://arxiv.org/abs/1904.10015}{{\ttfamily
  1904.10015}}].

\bibitem{Bandyopadhyay:2020ufc}
P.~Bandyopadhyay, M.~Mitra and A.~Roy, \emph{{Relativistic freeze-in with
  scalar dark matter in a gauged $B − L$ model and electroweak symmetry
  breaking}}, \href{https://doi.org/10.1007/JHEP05(2021)150}{\emph{JHEP}
  {\bfseries 05} (2021) 150}
  [\href{https://arxiv.org/abs/2012.07142}{{\ttfamily 2012.07142}}].

\bibitem{Minkowski:1977sc}
P.~Minkowski, \emph{{$\mu \to e\gamma$ at a Rate of One Out of $10^{9}$ Muon
  Decays?}}, \href{https://doi.org/10.1016/0370-2693(77)90435-X}{\emph{Phys.
  Lett. B} {\bfseries 67} (1977) 421}.

\bibitem{GellMann:1980vs}
M.~Gell-Mann, P.~Ramond and R.~Slansky, \emph{{Complex Spinors and Unified
  Theories}}, {\emph{Conf. Proc. C} {\bfseries 790927} (1979) 315}
  [\href{https://arxiv.org/abs/1306.4669}{{\ttfamily 1306.4669}}].

\bibitem{Mohapatra:1979ia}
R.~N. Mohapatra and G.~Senjanovic, \emph{{Neutrino Mass and Spontaneous Parity
  Nonconservation}},
  \href{https://doi.org/10.1103/PhysRevLett.44.912}{\emph{Phys. Rev. Lett.}
  {\bfseries 44} (1980) 912}.

\bibitem{10.1143/PTP.64.1103}
T.~Yanagida, \emph{{Horizontal Symmetry and Masses of Neutrinos}},
  \href{https://doi.org/10.1143/PTP.64.1103}{\emph{Progress of Theoretical
  Physics} {\bfseries 64} (1980) 1103}
  [\href{https://arxiv.org/abs/https://academic.oup.com/ptp/article-pdf/64/3/1103/5394376/64-3-1103.pdf}{{\ttfamily
  https://academic.oup.com/ptp/article-pdf/64/3/1103/5394376/64-3-1103.pdf}}].

\bibitem{Schechter:1980gr}
J.~Schechter and J.~W.~F. Valle, \emph{{Neutrino Masses in SU(2) x U(1)
  Theories}}, \href{https://doi.org/10.1103/PhysRevD.22.2227}{\emph{Phys. Rev.
  D} {\bfseries 22} (1980) 2227}.

\bibitem{Schechter:1981cv}
J.~Schechter and J.~W.~F. Valle, \emph{{Neutrino Decay and Spontaneous
  Violation of Lepton Number}},
  \href{https://doi.org/10.1103/PhysRevD.25.774}{\emph{Phys. Rev. D} {\bfseries
  25} (1982) 774}.

\bibitem{Brdar:2019iem}
V.~Brdar, A.~J. Helmboldt, S.~Iwamoto and K.~Schmitz, \emph{{Type-I Seesaw as
  the Common Origin of Neutrino Mass, Baryon Asymmetry, and the Electroweak
  Scale}}, \href{https://doi.org/10.1103/PhysRevD.100.075029}{\emph{Phys. Rev.
  D} {\bfseries 100} (2019) 075029}
  [\href{https://arxiv.org/abs/1905.12634}{{\ttfamily 1905.12634}}].

\bibitem{Fukuda:1998mi}
{\scshape Super-Kamiokande} collaboration, \emph{{Evidence for oscillation of
  atmospheric neutrinos}},
  \href{https://doi.org/10.1103/PhysRevLett.81.1562}{\emph{Phys. Rev. Lett.}
  {\bfseries 81} (1998) 1562}
  [\href{https://arxiv.org/abs/hep-ex/9807003}{{\ttfamily hep-ex/9807003}}].

\bibitem{Ahmad:2002jz}
{\scshape SNO} collaboration, \emph{{Direct evidence for neutrino flavor
  transformation from neutral current interactions in the Sudbury Neutrino
  Observatory}},
  \href{https://doi.org/10.1103/PhysRevLett.89.011301}{\emph{Phys. Rev. Lett.}
  {\bfseries 89} (2002) 011301}
  [\href{https://arxiv.org/abs/nucl-ex/0204008}{{\ttfamily nucl-ex/0204008}}].

\bibitem{Fogli:2006fw}
G.~L. Fogli, E.~Lisi, A.~Marrone, A.~Palazzo and A.~M. Rotunno, \emph{{Neutrino
  masses and neutrino mixing}},
  \href{https://doi.org/10.1016/j.nuclphysbps.2006.02.001}{\emph{Nucl. Phys. B
  Proc. Suppl.} {\bfseries 155} (2006) 5}.

\bibitem{Ahn:2002up}
{\scshape K2K} collaboration, \emph{{Indications of neutrino oscillation in a
  250 km long baseline experiment}},
  \href{https://doi.org/10.1103/PhysRevLett.90.041801}{\emph{Phys. Rev. Lett.}
  {\bfseries 90} (2003) 041801}
  [\href{https://arxiv.org/abs/hep-ex/0212007}{{\ttfamily hep-ex/0212007}}].

\bibitem{Fukugita:1986hr}
M.~Fukugita and T.~Yanagida, \emph{{Baryogenesis Without Grand Unification}},
  \href{https://doi.org/10.1016/0370-2693(86)91126-3}{\emph{Phys. Lett. B}
  {\bfseries 174} (1986) 45}.

\bibitem{Plumacher:1996kc}
M.~Plumacher, \emph{{Baryogenesis and lepton number violation}},
  \href{https://doi.org/10.1007/s002880050418}{\emph{Z. Phys. C} {\bfseries 74}
  (1997) 549} [\href{https://arxiv.org/abs/hep-ph/9604229}{{\ttfamily
  hep-ph/9604229}}].

\bibitem{Riotto:1999yt}
A.~Riotto and M.~Trodden, \emph{{Recent progress in baryogenesis}},
  \href{https://doi.org/10.1146/annurev.nucl.49.1.35}{\emph{Ann. Rev. Nucl.
  Part. Sci.} {\bfseries 49} (1999) 35}
  [\href{https://arxiv.org/abs/hep-ph/9901362}{{\ttfamily hep-ph/9901362}}].

\bibitem{Dine:2003ax}
M.~Dine and A.~Kusenko, \emph{{The Origin of the matter - antimatter
  asymmetry}}, \href{https://doi.org/10.1103/RevModPhys.76.1}{\emph{Rev. Mod.
  Phys.} {\bfseries 76} (2003) 1}
  [\href{https://arxiv.org/abs/hep-ph/0303065}{{\ttfamily hep-ph/0303065}}].

\bibitem{DuttaBanik:2020vfr}
A.~Dutta~Banik, R.~Roshan and A.~Sil, \emph{{Neutrino mass and asymmetric dark
  matter: study with inert Higgs doublet and high scale validity}},
  \href{https://doi.org/10.1088/1475-7516/2021/03/037}{\emph{JCAP} {\bfseries
  03} (2021) 037} [\href{https://arxiv.org/abs/2011.04371}{{\ttfamily
  2011.04371}}].

\bibitem{Datta:2021gyi}
A.~Datta, R.~Roshan and A.~Sil, \emph{{Scalar Triplet Flavor Leptogenesis with
  Dark Matter}},  \href{https://arxiv.org/abs/2110.03914}{{\ttfamily
  2110.03914}}.

\bibitem{Escudero:2018fwn}
M.~Escudero, S.~J. Witte and N.~Rius, \emph{{The dispirited case of gauged
  U(1)$_{B-L}$ dark matter}},
  \href{https://doi.org/10.1007/JHEP08(2018)190}{\emph{JHEP} {\bfseries 08}
  (2018) 190} [\href{https://arxiv.org/abs/1806.02823}{{\ttfamily
  1806.02823}}].

\bibitem{Basak:2013cga}
T.~Basak and T.~Mondal, \emph{{Constraining Minimal $U(1)_{B-L}$ model from
  Dark Matter Observations}},
  \href{https://doi.org/10.1103/PhysRevD.89.063527}{\emph{Phys. Rev. D}
  {\bfseries 89} (2014) 063527}
  [\href{https://arxiv.org/abs/1308.0023}{{\ttfamily 1308.0023}}].

\bibitem{Okada:2010wd}
N.~Okada and O.~Seto, \emph{{Higgs portal dark matter in the minimal gauged
  $U(1)_{B-L}$ model}},
  \href{https://doi.org/10.1103/PhysRevD.82.023507}{\emph{Phys. Rev. D}
  {\bfseries 82} (2010) 023507}
  [\href{https://arxiv.org/abs/1002.2525}{{\ttfamily 1002.2525}}].

\bibitem{Khalil:2008kp}
S.~Khalil and O.~Seto, \emph{{Sterile neutrino dark matter in B - L extension
  of the standard model and galactic 511-keV line}},
  \href{https://doi.org/10.1088/1475-7516/2008/10/024}{\emph{JCAP} {\bfseries
  10} (2008) 024} [\href{https://arxiv.org/abs/0804.0336}{{\ttfamily
  0804.0336}}].

\bibitem{Seto:2020udg}
O.~Seto and T.~Shimomura, \emph{{Signal from sterile neutrino dark matter in
  extra $U(1)$ model at direct detection experiment}},
  \href{https://doi.org/10.1016/j.physletb.2020.135880}{\emph{Phys. Lett. B}
  {\bfseries 811} (2020) 135880}
  [\href{https://arxiv.org/abs/2007.14605}{{\ttfamily 2007.14605}}].

\bibitem{Decant:2021mhj}
Q.~Decant, J.~Heisig, D.~C. Hooper and L.~Lopez-Honorez, \emph{{Lyman-$\alpha$
  constraints on freeze-in and superWIMPs}},
  \href{https://arxiv.org/abs/2111.09321}{{\ttfamily 2111.09321}}.

\bibitem{Montero:2007cd}
J.~C. Montero and V.~Pleitez, \emph{{Gauging U(1) symmetries and the number of
  right-handed neutrinos}},
  \href{https://doi.org/10.1016/j.physletb.2009.03.065}{\emph{Phys. Lett. B}
  {\bfseries 675} (2009) 64} [\href{https://arxiv.org/abs/0706.0473}{{\ttfamily
  0706.0473}}].

\bibitem{Hook:2010tw}
A.~Hook, E.~Izaguirre and J.~G. Wacker, \emph{{Model Independent Bounds on
  Kinetic Mixing}}, \href{https://doi.org/10.1155/2011/859762}{\emph{Adv. High
  Energy Phys.} {\bfseries 2011} (2011) 859762}
  [\href{https://arxiv.org/abs/1006.0973}{{\ttfamily 1006.0973}}].

\bibitem{deFlorian:2016spz}
{\scshape LHC Higgs Cross Section Working Group} collaboration, \emph{{Handbook
  of LHC Higgs Cross Sections: 4. Deciphering the Nature of the Higgs Sector}},
   \href{https://arxiv.org/abs/1610.07922}{{\ttfamily 1610.07922}}.

\bibitem{PhysRevD.45.2933}
M.~E. Carrington, \emph{Effective potential at finite temperature in the
  standard model}, \href{https://doi.org/10.1103/PhysRevD.45.2933}{\emph{Phys.
  Rev. D} {\bfseries 45} (1992) 2933}.

\bibitem{Giudice:2003jh}
G.~F. Giudice, A.~Notari, M.~Raidal, A.~Riotto and A.~Strumia, \emph{{Towards a
  complete theory of thermal leptogenesis in the SM and MSSM}},
  \href{https://doi.org/10.1016/j.nuclphysb.2004.02.019}{\emph{Nucl. Phys. B}
  {\bfseries 685} (2004) 89}
  [\href{https://arxiv.org/abs/hep-ph/0310123}{{\ttfamily hep-ph/0310123}}].

\bibitem{Comelli:1996vm}
D.~Comelli and J.~R. Espinosa, \emph{{Bosonic thermal masses in
  supersymmetry}}, \href{https://doi.org/10.1103/PhysRevD.55.6253}{\emph{Phys.
  Rev. D} {\bfseries 55} (1997) 6253}
  [\href{https://arxiv.org/abs/hep-ph/9606438}{{\ttfamily hep-ph/9606438}}].

\bibitem{Robens:2016xkb}
T.~Robens and T.~Stefaniak, \emph{{LHC Benchmark Scenarios for the Real Higgs
  Singlet Extension of the Standard Model}},
  \href{https://doi.org/10.1140/epjc/s10052-016-4115-8}{\emph{Eur. Phys. J. C}
  {\bfseries 76} (2016) 268}
  [\href{https://arxiv.org/abs/1601.07880}{{\ttfamily 1601.07880}}].

\bibitem{Carena:2004xs}
M.~Carena, A.~Daleo, B.~A. Dobrescu and T.~M.~P. Tait, \emph{{$Z^\prime$ gauge
  bosons at the Tevatron}},
  \href{https://doi.org/10.1103/PhysRevD.70.093009}{\emph{Phys. Rev. D}
  {\bfseries 70} (2004) 093009}
  [\href{https://arxiv.org/abs/hep-ph/0408098}{{\ttfamily hep-ph/0408098}}].

\bibitem{Cacciapaglia:2006pk}
G.~Cacciapaglia, C.~Csaki, G.~Marandella and A.~Strumia, \emph{{The Minimal Set
  of Electroweak Precision Parameters}},
  \href{https://doi.org/10.1103/PhysRevD.74.033011}{\emph{Phys. Rev. D}
  {\bfseries 74} (2006) 033011}
  [\href{https://arxiv.org/abs/hep-ph/0604111}{{\ttfamily hep-ph/0604111}}].

\bibitem{Das:2021esm}
A.~Das, P.~S.~B. Dev, Y.~Hosotani and S.~Mandal, \emph{{Probing the minimal
  $U(1)_X$ model at future electron-positron colliders via the fermion
  pair-production channel}},
  \href{https://arxiv.org/abs/2104.10902}{{\ttfamily 2104.10902}}.

\bibitem{ATLAS:2020kdi}
{\scshape ATLAS} collaboration, \emph{{Combination of searches for invisible
  Higgs boson decays with the ATLAS experiment}}, .

\bibitem{Bian:2021vmi}
L.~Bian, X.~Liu and K.-P. Xie, \emph{{Probing superheavy dark matter with
  gravitational waves}},  \href{https://arxiv.org/abs/2107.13112}{{\ttfamily
  2107.13112}}.

\bibitem{McLaughlin:2013ira}
M.~A. McLaughlin, \emph{{The North American Nanohertz Observatory for
  Gravitational Waves}},
  \href{https://doi.org/10.1088/0264-9381/30/22/224008}{\emph{Class. Quant.
  Grav.} {\bfseries 30} (2013) 224008}
  [\href{https://arxiv.org/abs/1310.0758}{{\ttfamily 1310.0758}}].

\bibitem{Shannon:2015ect}
R.~M. Shannon et~al., \emph{{Gravitational waves from binary supermassive black
  holes missing in pulsar observations}},
  \href{https://doi.org/10.1126/science.aab1910}{\emph{Science} {\bfseries 349}
  (2015) 1522} [\href{https://arxiv.org/abs/1509.07320}{{\ttfamily
  1509.07320}}].

\bibitem{Kramer:2013kea}
M.~Kramer and D.~J. Champion, \emph{{The European Pulsar Timing Array and the
  Large European Array for Pulsars}},
  \href{https://doi.org/10.1088/0264-9381/30/22/224009}{\emph{Class. Quant.
  Grav.} {\bfseries 30} (2013) 224009}.

\bibitem{Hobbs:2009yy}
G.~Hobbs et~al., \emph{{The international pulsar timing array project: using
  pulsars as a gravitational wave detector}},
  \href{https://doi.org/10.1088/0264-9381/27/8/084013}{\emph{Class. Quant.
  Grav.} {\bfseries 27} (2010) 084013}
  [\href{https://arxiv.org/abs/0911.5206}{{\ttfamily 0911.5206}}].

\bibitem{LISA:2017pwj}
{\scshape LISA} collaboration, \emph{{Laser Interferometer Space Antenna}},
  \href{https://arxiv.org/abs/1702.00786}{{\ttfamily 1702.00786}}.

\bibitem{LIGOScientific:2019vic}
{\scshape LIGO Scientific, Virgo} collaboration, \emph{{Search for the
  isotropic stochastic background using data from Advanced
  LIGO\textquoteright{}s second observing run}},
  \href{https://doi.org/10.1103/PhysRevD.100.061101}{\emph{Phys. Rev. D}
  {\bfseries 100} (2019) 061101}
  [\href{https://arxiv.org/abs/1903.02886}{{\ttfamily 1903.02886}}].

\end{thebibliography}\endgroup


\end{document}